\documentclass[12pt]{article}
\pdfoutput=1
\usepackage[colorlinks,linkcolor=Blue,citecolor=Blue,bookmarks,bookmarksnumbered]{hyperref}
\usepackage[scaled=0.85]{helvet}
\usepackage{amsmath,amssymb,accents,mathrsfs}
\usepackage{XoohmE}
\usepackage{graphicx,color}
\usepackage{booktabs}
\usepackage{multirow}
\usepackage{placeins}
\usepackage{amsmath}
\usepackage{cases}

\definecolor{Green}  {rgb}{0.10,0.70,0.10} %  1
\definecolor{Orange} {rgb}{1.00,0.50,0.15} %  2
\definecolor{Red}    {rgb}{0.90,0.00,0.12} %  3
\definecolor{Purple} {rgb}{0.50,0.25,0.55} %  4
\definecolor{Turque} {rgb}{0.00,0.65,0.85} %  5
\definecolor{Blue}   {rgb}{0.00,0.00,1.00} %  6
\definecolor{Magenta}{rgb}{1.00,0.00,1.00} %  7
\definecolor{Gold}   {rgb}{1.00,0.75,0.25} %  8
\definecolor{Seaweed}{rgb}{0.01,0.24,0.09} %  9
\definecolor{Brown}  {rgb}{0.43,0.26,0.32} % 10
\definecolor{grey1}  {rgb}{0.20,0.20,0.20} % 11
\definecolor{grey2}  {rgb}{0.40,0.40,0.40} % 12
\definecolor{grey3}  {rgb}{0.60,0.60,0.60} % 13
\definecolor{grey4}  {rgb}{0.80,0.80,0.80} % 14
\definecolor{grey5}  {rgb}{0.90,0.90,0.90} % 15
\def\C#1#2{{\ifcase#1\or%Greg's color scheme
             \color{Green}\or \color{Orange}\or \color{Red}\or
              \color{Purple}\or \color{Turque}\or \color{Blue}\or
               \color{Magenta}\or \color{Gold}\or \color{Seaweed}\or
                \color{Brown}\or\color{grey1}\or\color{grey2}\or
                 \color{grey3}\else\color{grey4}\fi#2}}

\definecolor{Slate} {rgb}{0.00,0.45,0.55}

\def\be{\begin{equation}}
\def\ee{\end{equation}}
\newcommand{\bea}{\begin{eqnarray}}
\newcommand{\eea}{\end{eqnarray}}
\newcommand{\ena}{\end{eqnarray}}

\def\pp{{\mathchoice
          {%displaystyle
              \kern 1pt%
              \raise 1pt
              \vbox{\hrule width5pt height0.4pt depth0pt
                    \kern -2pt
                    \hbox{\kern 2.3pt
                          \vrule width0.4pt height6pt depth0pt
                          }
                    \kern -2pt
                    \hrule width5pt height0.4pt depth0pt}%
                    \kern 1pt
           }
            {%textstyle
              \kern 1pt%
              \raise 1pt
              \vbox{\hrule width4.3pt height0.4pt depth0pt
                    \kern -1.8pt
                    \hbox{\kern 1.95pt
                          \vrule width0.4pt height5.4pt depth0pt
                          }
                    \kern -1.8pt
                    \hrule width4.3pt height0.4pt depth0pt}%
                    \kern 1pt
            }
            {%scriptstyle
              \kern 0.5pt%
              \raise 1pt
              \vbox{\hrule width4.0pt height0.3pt depth0pt
                    \kern -1.9pt  %[e]=0.15pt
                    \hbox{\kern 1.85pt
                          \vrule width0.3pt height5.7pt depth0pt
                          }
                    \kern -1.9pt
                    \hrule width4.0pt height0.3pt depth0pt}%
                    \kern 0.5pt
            }
            {%scriptscriptstyle
              \kern 0.5pt%
              \raise 1pt
              \vbox{\hrule width3.6pt height0.3pt depth0pt
                    \kern -1.5pt
                    \hbox{\kern 1.65pt
                          \vrule width0.3pt height4.5pt depth0pt
                          }
                    \kern -1.5pt
                    \hrule width3.6pt height0.3pt depth0pt}%
                    \kern 0.5pt%}
            }
        }}

\def\mm{{\mathchoice
                       {%displaystyle
                             \kern 1pt
               \raise 1pt    \vbox{\hrule width5pt height0.4pt depth0pt
                                  \kern 2pt
                                  \hrule width5pt height0.4pt depth0pt}
                             \kern 1pt}
                       {%textstyle
                            \kern 1pt
               \raise 1pt \vbox{\hrule width4.3pt height0.4pt depth0pt
                                  \kern 1.8pt
                                  \hrule width4.3pt height0.4pt depth0pt}
                             \kern 1pt}
                       {%scriptstyle
                            \kern 0.5pt
               \raise 1pt
                            \vbox{\hrule width4.0pt height0.3pt depth0pt
                                  \kern 1.9pt
                                  \hrule width4.0pt height0.3pt depth0pt}
                            \kern 1pt}
                       {%scriptscriptstyle
                           \kern 0.5pt
             \raise 1pt  \vbox{\hrule width3.6pt height0.3pt depth0pt
                                  \kern 1.5pt
                                  \hrule width3.6pt height0.3pt depth0pt}
                           \kern 0.5pt}
                       }}

\def\ad{{\kern0.5pt
                   \alpha \kern-5.05pt \raise5.8pt\hbox{$\textstyle.$}\kern
0.5pt}}

\def\bd{{\kern0.5pt
                   \beta \kern-5.05pt \raise5.8pt\hbox{$\textstyle.$}\kern
0.5pt}}

\def\qd{{\kern0.5pt
                   q \kern-5.05pt \raise5.8pt\hbox{$\textstyle.$}\kern
0.5pt}}
\def\Dot#1{{\kern0.5pt
     {#1} \kern-5.05pt \raise5.8pt\hbox{$\textstyle.$}\kern
0.5pt}}

\catcode`@=11
\def\un#1{\relax\ifmmode\@@underline#1\else
        $\@@underline{\hbox{#1}}$\relax\fi}
\catcode`@=12

\def\a{\alpha}
\def\b{\beta}
\def\c{\chi}
\def\d{\delta}

\def\g{\gamma}

\def\i{\iota}
\def\j{\psi}
\def\k{\kappa}
\def\l{\lambda}

\def\n{\nu}
\def\o{\omega}

\def\r{\rho}

\def\t{\tau}

\def\D{\Delta}

\def\L{\Lambda}
\def\O{\Omega}
\def\P{\Pi}

\def\S{\Sigma}

\def\ca{{\cal A}}

\def\cf{{\cal F}}

\def\dslash{\not{\hbox{\kern-2pt $\partial$}}}
\def\Dslash{\not{\hbox{\kern-4pt $D$}}}
\def\pslash{\not{\hbox{\kern-2.3pt $p$}}}
 \newtoks\slashfraction
 \slashfraction={.13}
 \def\slash#1{\setbox0\hbox{$ #1 $}
 \setbox0\hbox to \the\slashfraction\wd0{\hss \box0}/\box0 }
 
\def\kcr{{\hbox{\ro \char'170}}}                % right-handed rope
\def\ktl{{\hbox{\ro \char'170}}}        % top end for left-handed rope
\def\ktr{{\hbox{\ro \char'170}}}        % " right
\def\kbl{{\hbox{\ro \char'170}}}        % " bottom left
\def\kbr{{\hbox{\ro \char'170}}}        % " right
\def\plpl{\raise-2pt\hbox{$\raise3pt\hbox{$_+$}\hskip-6.67pt\raise0.0pt
\hbox{$^+$}\hskip 0.01pt$}}
\def\mimi{\raise-2pt\hbox{$\raise3pt\hbox{$_-$}\hskip-6.67pt\raise0.0pt
\hbox{$^-$}\hskip 0.01pt$}} 

\def\bo{{\raise.15ex\hbox{\large$\Box$}}}               % D'Alembertian
\def\pa{\partial}                                       % curly d
\def\TH{{\raise.2ex\hbox{$\displaystyle \bigodot$}\mskip-4.7mu \llap H \;}}
\def\face{{\raise.2ex\hbox{$\displaystyle \bigodot$}\mskip-2.2mu \llap {$\ddot
        \smile$}}}                                      % happy face
\def\dt#1{\on{\hbox{\bf .}}{#1}}                % (big) dot over
\def\Dot#1{\dt{#1}}

\def\Bar#1{\overline{#1}}                       % big bar
\def\leftrightarrowfill{$\mathsurround=0pt \mathord\leftarrow \mkern-6mu
        \cleaders\hbox{$\mkern-2mu \mathord- \mkern-2mu$}\hfill
        \mkern-6mu \mathord\rightarrow$}
\def\dvec#1{\vbox{\ialign{##\crcr
        \leftrightarrowfill\crcr\noalign{\kern-1pt\nointerlineskip}
        $\hfil\displaystyle{#1}\hfil$\crcr}}}           % <--> accent
\def\dt#1{{\buildrel {\hbox{\LARGE .}} \over {#1}}}     % dot-over for sp/sb
\def\sfrac#1#2{{\vphantom1\smash{\lower.5ex\hbox{\small$#1$}}\over
        \vphantom1\smash{\raise.4ex\hbox{\small$#2$}}}} % alternate fraction
\def\bfrac#1#2{{\vphantom1\smash{\lower.5ex\hbox{$#1$}}\over
        \vphantom1\smash{\raise.3ex\hbox{$#2$}}}}       % "
\def\afrac#1#2{{\vphantom1\smash{\lower.5ex\hbox{$#1$}}\over#2}}    % "
\def\pa{\partial}

\def\ad{{\dot{\alpha}}}
\def\bd{{\dot{\beta}}}

 \font\rOpe=cmsy10                        % Ersatz for the non-standard rope font
 \def\ktl{{\hbox{\rOpe\char'170}}}        % top end for left-handed rope
 \def\kbl{{\hbox{\rOpe\char'170}}}        % bottom end for left-handed rope
 \def\kcr{{\reflectbox{\rOpe\char'170}}}        % right-handed rope
 \def\ktr{{\reflectbox{\rOpe\char'170}}}        % top end for right-handed rope
 \def\kbr{{\reflectbox{\rOpe\char'170}}}        % bottom end for right-handed rope
 \def\Border{\vbox{\hsize0pt% braided border
        \setlength{\unitlength}{1mm}
        \newcount\xco
        \newcount\yco
        \xco=-21
        \yco=12
        \begin{picture}(0,0)(-7.5,0)
        \put(\xco,\yco){$\ktl$}
        \advance\yco by-1
        {\loop
        \put(\xco,\yco){$\kcr$}
        \advance\yco by-2
        \ifnum\yco>-240
        \repeat
        \put(\xco,\yco){$\kbl$}}
        \xco=170
        \yco=12
        \put(\xco,\yco){$\ktr$}
        \advance\yco by-1
        {\loop
        \put(\xco,\yco){$\kcr$}
        \advance\yco by-2
        \ifnum\yco>-240
        \repeat
        \put(\xco,\yco){$\kbr$}}
        \put(-19.5,13){\scalebox{.6065}{%
         University of Maryland Center for String and Particle  Theory \&\ Physics Department%
        |University of Maryland Center for String and Particle  Theory \&\ Physics Department}}
        \put(-19.5,-241.5){\scalebox{.5835}{%
         ****University of Maryland * Center for String and
         Particle  Theory* Physics Department****University of Maryland *Center
        for String and Particle  Theory* Physics Department}}
        \end{picture}
        \par\vskip-8mm}}
\definecolor{UMred}{rgb}{.9,.05,.2}
\definecolor{HUblue}{rgb}{.0,.3,.7}
 \def\UMbanner{\vbox{\hsize0pt% The "UM" banner
        \setlength{\unitlength}{.4mm}
        \thicklines  
        \begin{picture}(0,0)(-30,-10)
        \put(165,2){\line(1,0){4}}
        \put(170,2){\line(1,0){4}}
        \put(180,2){\line(1,0){4}}
        \put(175,-14){\line(1,0){4}}
        \put(180,-14){\line(1,0){4}}
        \put(185,-14){\line(1,0){4}}
        \put(169,-14){\line(0,1){16}}
        \put(170,-14){\line(0,1){16}}
        \put(179,-14){\line(0,1){16}}
        \put(180,-14){\line(0,1){16}}
        \put(184,-14){\line(0,1){16}}
        \put(185,-14){\line(0,1){16}}
        \put(169,2){\oval(8,32)[bl]}
        \put(170,2){\oval(8,32)[br]}
        \put(179,-14){\oval(8,32)[tl]}
        \put(185,-14){\oval(8,32)[tr]}
        \end{picture}
        \par\vskip-6.5mm
        \thicklines}}

\definecolor{Red}    {rgb}{0.90,0.00,0.12} %  1
\definecolor{Blue}   {rgb}{0.00,0.00,1.00} %  2
\definecolor{Green}  {rgb}{0.10,0.70,0.10} %  3
\definecolor{Turque} {rgb}{0.00,0.65,0.85} %  4
\definecolor{Orange} {rgb}{1.00,0.50,0.15} %  5
\definecolor{Magenta}{rgb}{1.00,0.00,1.00} %  6
\definecolor{Gold}   {rgb}{1.00,0.75,0.25} %  7
\definecolor{Seaweed}{rgb}{0.01,0.24,0.09} %  8
\definecolor{Purple} {rgb}{0.50,0.25,0.55} %  9
\definecolor{Brown}  {rgb}{0.43,0.26,0.32} % 10
\definecolor{grey1}  {rgb}{0.20,0.20,0.20} % 11
\definecolor{grey2}  {rgb}{0.40,0.40,0.40} % 12
\definecolor{grey3}  {rgb}{0.60,0.60,0.60} % 13
\definecolor{grey4}  {rgb}{0.80,0.80,0.80} % 14
\definecolor{grey5}  {rgb}{0.90,0.90,0.90} % 15
\def\C#1#2{{\ifcase#1\or%TH color scheme
             \color{Red}\or \color{Green}\or \color{Blue}\or\
              \color{Turque}\or \color{Orange}\or \color{Magenta}\or 
               \color{Gold}\or \color{Seaweed}\or \color{Purple}\or
                \color{Brown}\or\color{grey1}\or\color{grey2}\or
                 \color{grey3}\else\color{grey4}\fi#2}}

\definecolor{Slate} {rgb}{0.00,0.45,0.55}

\newdimen\parshift\parshift=\parindent
\catcode`@=11
 \long\def\@footnotetext#1{\insert\footins{\reset@font\footnotesize
           \interlinepenalty\interfootnotelinepenalty\splittopskip%
            \footnotesep\splitmaxdepth\dp\strutbox\floatingpenalty\@MM%
             \hsize\columnwidth\addtolength{\hsize}{-2\parindent}
              \@parboxrestore\protected@edef\@currentlabel%
              {\csname p@footnote\endcsname\@thefnmark}%
                \color@begingroup%
                 \@makefntext{\rule\z@\footnotesep\ignorespaces#1%
                  \@finalstrut\strutbox}%
                \color@endgroup}}
 \long\def\@makefntext#1{\hglue\parshift%
           \vbox{\noindent\baselineskip=11pt plus.5pt minus.5pt\hb@xt@0em{\hss\@makefnmark\kern1pt}#1}}
\catcode`@=12

\usepackage[retainorgcmds]{IEEEtrantools}          
\def\Ibea{\begin{IEEEeqnarray*}}
\def\Ieea{\end{IEEEeqnarray*}}
\def\n{\IEEEyesnumber}
\def\sn{\IEEEyessubnumber}

\def\ad{{\dot{\alpha}}}  %%%% Dotted Greek Indices %%%%
\def\bd{{\dot{\beta}}}
\def\gd{{\dot{\gamma}}}

\def\D{{\rm D}}         %%%% Operators / Equations %%%%
\def\Dd{{\bar{\rm D}}}
\def\pa{\partial}

\newskip\humongous \humongous=0pt plus 1000pt minus 1000pt
\def\caja{\mathsurround=0pt}

\newif\ifdtup
\def\panorama{\global\dtuptrue \openup2\jot \caja
        \everycr{\noalign{\ifdtup \global\dtupfalse
        \vskip-\lineskiplimit \vskip\normallineskiplimit
        \else \penalty\interdisplaylinepenalty \fi}}}

\makeatletter
\def\section{\@startsection{section}{1}{\z@}
        {3ex plus-1ex minus-.2ex}{1pt plus1pt}{\large\sf\bfseries\boldmath}}
\def%%%%%%%%%%%%%%%%%%%%%%%%%%%%%%%%%
\subsection{\@startsection{subsection}{2}{\z@}
         {1.5ex plus-1ex minus-.2ex}{0.01pt plus1pt}{\sf\slshape}}
\def\subsubsection{\@startsection{subsubsection}{3}{\z@}
          {1.5ex plus-1ex minus-.2ex}{0.01pt plus0.2pt}{\sf\boldmath}}
\def\paragraph{\@startsection{paragraph}{4}{\z@}
           {.75ex \@plus.5ex \@minus.2ex}{-2mm}{\sf\bfseries\boldmath}}
\makeatother

 \allowdisplaybreaks
 \seceq

\begin{document}

\thispagestyle{empty}
\vbox{\Border\UMbanner}
\noindent{\small
%\today
\hfill{PP--017-24 {~} \\ % un-comment-out and specify when done}  
$~~~~~~~~~~~~~~~~~~~~~~~~~~~~~~~~~~~~~~~~~~~~~~~~~~~~~~~~~~~~$
$~~~~~~~~~~~~~~~~~~~~\,~~~~~~~~~~~~~~~~~~~~~~~~~\,~~~~~~~~~~~~~~~~$
 {HET-1715}
}
\vspace*{8mm}
\begin{center}
{\large \bf
From Diophantus to Supergravity and 
\\ [2pt]
massless higher spin multiplets
}   \\   [12mm]
{\large {
S.\ James Gates, Jr.,\footnote{gatess@wam.umd.edu}$^{a, \, b}$
and Konstantinos Koutrolikos\footnote{kkoutrolikos@physics.muni.cz}$^{c}$
}}
\\*[10mm]
\emph{
\centering
$^a$Center for String and Particle Theory-Dept.\ of Physics,
University of Maryland, \\[-2pt]
4150 Campus Dr., College Park, MD 20472,  USA
\\[11pt]         
$^{b}$Department of Physics, Brown University,
\\[1pt]
Box 1843, 182 Hope Street, Barus \& Holley 545,
Providence, RI 02912, USA 
\\[11pt]
and
\\[11pt] 
$^c$ Institute for Theoretical Physics and Astrophysics,
\\[1pt]
Masaryk University,
\\[1pt]
611 37 Brno, Czech Republic
}
 $$~~$$
  $$~~$$
 \\*[-7mm]
{ ABSTRACT}\\[4mm]
\parbox{142mm}{\parindent=2pc\indent\baselineskip=14pt plus1pt
We present a new method of deriving the off-shell spectrum of 
supergravity and massless $4D,~\mathcal{N}=1$ higher spin multiplets 
without the need of an action and based on a set of natural requirements:
(a.) existence of an underlying superspace description, (b.) an economical 
description of free, massless, higher spins and (c.) equal numbers of bosonic 
and fermionic degrees of freedom. We prove that for any theory that respects the above,
the fermionic auxiliary components come in pairs and are gauge invariant and there are
two types of bosonic auxiliary components. Type (1) are pairs of a $(2,0)$-tensor
with real or imaginary $(1,1)$-tensor with non-trivial gauge transformations.
Type (2) are singlets and gauge invariant. The outcome is a set of Diophantine
equations, the solutions of which determine the off-shell spectrum of supergravity and 
massless higher spin multiplets. This approach provides (\emph{i}) a classification of
the irreducible, supersymmetric, representations of arbitrary spin and (\emph{ii}) a very clean and intuitive
explanation to why some of these theories have more than one formulations (e.g. the supergravity multiplet) and others do not.
 }
 \end{center}
 $$~~$$
$$~~$$
\vfill
\noindent PACS: 11.30.Pb, 12.60.Jv\\
Keywords: supersymmetry, off-shell supermultiplets
\vfill
\clearpage

\section{Introduction}

The study of higher spin theories is very well motivated these days in part 
due to their increasing relevance originating from string theory and holography. 
In string theory, one can argue that many of the attractive UV properties of 
scattering amplitudes calculated perturbatively originate from the transmission 
of higher spin particles at the corresponding exchange. Also, there is the old 
conjecture that the massless phase of the theory emerging at some ultra-high
 energy regime (i.e. the ``tensionaless string''), will be governed by some sort 
 of a higher spin gauge theory.  For holography, the appearance of higher 
 spins is slightly more subtle. Interacting higher spin theories are very difficult 
 to construct, but there has been some success and all of them have the same two features
 in common. The existence of a massless spin 2 state and a non-flat 
 background (usually AdS).  These two properties, make higher spins the 
 perfect laboratory for investigating holography.

In many of the above described theories, supersymmetry is a vital ingredient and 
therefore it is natural to ask for the description of higher spins in the 
presence of supersymmetry. Higher spin irreducible representations (irreps) 
of the supersymmetric extension of the Poincar\'{e} symmetry group have 
been studied in detail in 4D where the meaning of spin and mass is clear. 
On the one hand there is the, group theoretic, on-shell approach which 
provides the set of conditions we must impose on tensors in order to 
describe irreducible supermultiplets. This is very important because 
every theory must respect these conditions at the zero coupling limit 
and properly deform, without negating, them when interactions are 
restored. One of the well known results of this approach is the on-shell 
spectrum of the theory, which is related with the decomposition of the 
irreps of the super-Poincar\'{e} group to irreps of the Poincar\'{e} group. 
For example, massless, $\mathcal{N}=1$ 
supermultiplets of superspin $Y$ include one spin $j=Y$ 
and one spin $j=Y+1/2$. This is due to the fact that for massless theories 
only $1/4$ of the supersymmetry generators can be used for creation 
operators (ascending ladder operators) and each one can be used at 
most once. Therefore for $\mathcal{N}=1$, there is only one creation 
operator and therefore we have two states, the `vacuum' ($j=Y$) and 
its first excitation $(j$=$Y$+1/2$)$.

On the other hand, there is the off-shell description of the theory, where 
one is writing actions that provide equations of motion compatible with 
the previously mentioned on-shell conditions.
This is the first step towards the fully interacting theory (if it exists). 
Such constructions exist for massless $\mathcal{N}=1$ theories of arbitrary 
high spin \cite{K1,K2,G1,G2,G3} but not for massive theories\cite{G4,K3}. 
The main result of this approach is the off-shell spectrum of the theory, 
meaning the set of auxiliary components that supersymmetry introduces 
on top of the components required for the description of propagating higher spin fields. 
Although there has been a very detailed derivation \cite{G1,G2,G3} of the 
off-shell spectrum for massless theories, there is no intuitive argument
that leads directly to these results.

The purpose of this paper is to fill this gap and provide an
alternative derivation of the set of off-shell auxiliary components required by supersymmetry
without having to (a.) find the free higher spin superspace action first and then (b.)  
project to components to extract the off-shell spectrum of the theory.
We will show that there is a set of natural requirements 
that fix the auxiliary components completely for any spin and allow us to 
deduce the required gauge transformations that lead uniquely to the action 
and generate all the details.  In other words, we present a method that given 
the on-shell spectrum of massless $4D, \mathcal{N}=1$ theories (the spins we want to describe)
produces the off-shell completion, \emph{i.e.} the set of auxiliary fields required, their
gauge transformations, 
the set of superfields that generate these fields and give the superspace 
description of the theory and their superspace
gauge transformations. We hope this addition to the understanding of the
structure of the auxiliary components of supersymmetric theories can 
eventually help towards finding the superspace description of massive 
higher spin systems and their future development.

The method relies on a list of requirements, whose imposition extracts
the set of desired theories from the set of all possible 
theories.  As we are interested in supersymmetric theories, one of the 
requirements must provide this information. We do this by demanding 
the equality of the bosonic degrees of freedom (d.o.f) of the theory
with the fermionic d.o.f.
This is a natural requirement that all supersymmetric theories must respect off-shell. 
For future reference we will call this the \emph{Supersymmetry} requirement.

Furthermore, it is natural to have the property of supersymmetry manifest.
This would mean that there must exist a superspace description of the theory,
hence we demand that all the components in the off-shell spectrum of the theory
must be generated by superfields and perfectly organise into multiplets.
We will call this the \emph{Superspace} requirement.

Another key property these theories have, is that they describe
free, massless irreps of one spin $j=Y+1/2$ and one $j=Y$. This means that among the components of
the off-shell spectrum we must have (a.) the correct number and type of components we need
for the correct off-shell description of exactly one spin $j=Y+1/2$ and exactly one spin $j=Y$ 
together with the appropriate gauge transformations, (b.) no other propagating d.o.f . We will
call this requirement the \emph{Higher spin} requirement.
  
The last requirement we demand is that of \emph{Economy}. Although the 
previous three requirements define the type of theories we want to 
describe, they do not say anything about their complexity. So it is 
reasonable to ask for the most economical 
description of these systems. That means economy in terms of the 
number of superfields required, their rank as tensors and degrees 
of freedom.

In this paper we will show that by imposing the requirements of (i) \emph{Supersymmetry}, (ii) \emph{
Superspace}, (iii) \emph{Higher spin} and 
(iv) \emph{Economy} as explained above we can determine the 
off-shell structure of the theory and the superspace description. The appeal of this 
approach is that it is very intuitive and, with great satisfaction, gives 
the answers obtained by following the technical constructions
of \cite{G1,G2,G3}.

The organization of the paper is the following. In sections 2, 3 and 4 we will 
derive the consequences of the above mentioned requirements.
That will provide a classification of the various types of components and superfields
we must consider as well as the type of gauge transformations they must have. Also, we derive
a constraint on the number of components of each type, in the form of a Diophantine equation.
In sections 5, 6 and 7 we apply these results to the
vector multiplet ($Y=\tfrac{1}{2}$), the
matter-gravitino multiplet ($Y=1$) and the supergravity multiplet ($Y=3/2$).
We find that this method will correctly generate all known formulations for these supermultiplets
[one for the vector multiplet, two for the gravitino matter multiplet (de Wit-van Holten, Ogievetsky-Sokatchev)
and four for the supergravity (old-minimal, new-minimal, new-new-minimal and non-minimal)]. 
In section 8, we extend the construction
to the arbitrary half-integer case $Y=s+1/2$ and correctly generate the two different 
ways to construct the supermultiplet in agreement with \cite{K1,G2}. The arbitrary integer 
case $Y=s$ is being discussed in section 9 and we find that there is a unique 
answer in agreement with \cite{G3}.
%%%%%%%%%%%%%%%%%%%%%%%%%%%%%%%%%%%%%%
%%%%%%%%%%%%%%%%%%%%%%%%%%%%%%%%%%%%%%
%%%%% - Higher spin - %%%%%%%%%%%%%%%%
\section{On-shell data and the \emph{Higher spin} requirement}
Based on the group theoretic analysis of the irreducible representations of the 
super-Poincar\'{e} group, we know that on-shell they must describe two successive 
spins, one integer and one half-integer. The off-shell description of these free, massless,
higher spins is very well known \cite{hs3,hs4,hs1,hs2}. Very briefly, in order to
describe these higher spins irreps:
\begin{enumerate}
\item[$\a$)] For integer spin $j=s$, we must have two real bosonic components 
$h_{\a(s)\ad(s)}$, $h_{\a(s-2)\ad(s-2)}$ with independently symmetrized dotted 
and undotted indices
\footnote{The notation $A_{\a(k)\ad(l)}$ is a shorthand for $A_{\a_1\a_2...\a_{k}\ad_1\ad_{2}...\ad_{l}}$ which is
a $(k,l)$-tensor with $k$ independently\\ symmetrized undotted indices and $l$ independently symmetrized dotted indices.}
along with appropriate gauge transformations
\Ibea{l}\n
\delta_g h_{\a(s)\ad(s)}\sim\pa_{(\a_s(\ad_s}\zeta_{\a(s-1))\ad(s-1))}~,\sn\\
\delta_g h_{\a(s-2)\ad(s-2)}\sim\pa^{\b\bd}\zeta_{\b\a(s-2)\bd\ad(s-2)}~.\sn
\Ieea
It is straightforward to count the off-shell degrees of freedom for spin $s$, the answer is $s^2+2$
\footnote{It may be useful to remind the reader that a $(k,l)$-tensor $A_{\a(k)\ad(l)}$ in $4D$ has $2(k+1)(l+1)$ d.o.f.\\
However if $k=l$ then we can talk about reality. A real $(k,k)$-tensor $A_{\a(k)\ad(k)}$ has $(k+1)^2$ d.o.f.}
.
\item[$\b$)] For half-integer spin $j=s+1/2$, we must have three fermionic components 
$\psi_{\a(s+1)\ad(s)}$, $\psi_{\a(s)\ad(s-1)}$ and $\psi_{\a(s-1)\ad(s-2)}$ also with 
independently symmetrized indices along with gauge transformations
\Ibea{l}\n
\delta_g \psi_{\a(s+1)\ad(s)}\sim\pa_{(\a_{s+1}(\ad_s}\xi_{\a(s))\ad(s-1))}~,\sn\\
\delta_g \psi_{\a(s)\ad(s-1)}\sim\pa_{(\a_{s}}{}^{\bd}\bar{\xi}_{\a(s-1))\bd\ad(s-1)}~,\sn\\
\delta_g \psi_{\a(s-1)\ad(s-2)}\sim\pa^{\b\bd}\xi_{\b\a(s-1)\bd\ad(s-2)}~.\sn
\Ieea
The counting 
of the off-shell degrees of freedom for spin $s+1/2$ is $4s^2+4s+4$.
\end{enumerate}
Hence the off-shell spectrum of the higher spin supersymmetric theory must respect these
structures and must include exactly one copy of these fields for each spin
it describes on-shell.
Notice that the off-shell bosonic and fermionic degrees of freedom do not match. This 
is one way to realize that in supersymmetric theories we must introduce extra auxiliary
fields in order to balance the bosonic and fermionic d.o.f, as requested by our \emph{Supersymmetry}
requirement.
%%%%%%%%%%%%%%%%%%%%%%%%%%%%%%%%%%%%%%
%%%%%%%%%%%%%%%%%%%%%%%%%%%%%%%%%%%%%%
%%%%%%%% - Supersymmetry - %%%%%%%%%%%
\section{Demanding \emph{Supersymmetry}}
\subsection{Types of auxiliary fields}\label{type1}
The extra fields we have to add must not introduce any new propagating d.o.f, because the on-shell spectrum
of the theory has already been taken care by the higher spin fields above. Therofore, the auxiliary fields must
have no dynamics, hence they can be defined\footnote{After appropriate 
redefinitions to absorb} 
such that either they vanish on-shell or they are equivalent to zero up to gauge transformations.
This mean that their equations of motion give a unique 
solution\footnote{For simplicity we suppressed all indices}:
\bea
A=0~~\text{or}~~A=0+\text{pure gauge}~.
\eea
The first one $[A=0]$ means that the equations of motion for these auxiliary fields are 
algebraic and therefore they must appear in the lagrangian in an algebraic manner. 
There are two ways this can happen, either they must appear in pairs $(A,B)$ 
[$\mathcal{L}=\dots+AB$] with engineering dimensions $d_A+d_B=4$
\footnote{Keep in mind that the enginnering dimensions of bosonic fields is an integer number and for fermionic fields it is a half integer number.
We have not proved that, but it will become obvious in the next section.} or as singles
[$\mathcal{L}=\dots+A^2$] with engineering dimensions $d_A=2 $. Immediately we get that
the fermionic 
auxiliary fields in this category must always appear in pairs, while the bosonic ones can
appear as singles. Another important characteristic feature of these auxiliary 
fields is that all of them have to be gauge invariant ($\delta_g A=0$) since the right hand side of 
the solution is gauge invariant .

The second type of solution $[A=0+\text{pure gauge}]$ means the auxiliary fields have a non-trivial transformation $
\delta_g A_{...}\sim \pa^{..}_{..}\lambda_{...}$ with appropriate index contractions and 
symmetrizations to match the index structure of $A$. For this case, the equation of 
motion can be differential ($\mathcal{O}A=0$, where $\mathcal{O}$ is a differential
operator) but it must have a unique solution which respects the gauge transformation of $A$, in other words
the following must be true
\bea
\mathcal{O}\cdot\pa\cdot\lambda=0~~\text{(identically)}
\eea
thus putting $A$ in the same equivalence class as the zero solution. Assuming that 
$\mathcal{O}$ includes one derivative\footnote{Two or more derivatives will lead to 
dynamics which these fields must not have. For the same reason $A$ can not be a 
fermion because fermions have one derivative equations of motion.} we obtain
\bea
\pa\cdot\pa\cdot\lambda=0~~\text{(identically)}
\eea
where the dot $[\cdot]$ represent the various choices for contractions and symmetrization of 
the indices. We can go through the list of all options (all indices contracted, all indices 
symmetrizes, ...) but we quickly realize that in most of the cases the equation does not 
vanish identically. The way out is to realize that the identical nature must of the equation relay on the 
symmetries of the indices and since $\lambda$ must have symmetrized indices then 
the product $\pa\cdot\pa$ must generate an antisymmetrization. Such an identity exists
\bea
\pa_{\a}{}^{\ad}\pa_{\b\ad}\sim C_{\a\b}\Box,~~C_{\a\b}=-C_{\b\a}~.
\eea
Therefore, we conclude that $A$ must be of the form $A_{\a(k+1)\ad}$ and its equation of 
motion and gauge transformation are
\bea
\pa_{(\a_{k+2}}{}^{\ad}A_{\a(k+1))\ad}=0~,~\delta_g A_{\a(k+1)\ad}\sim\pa_{
(\a_{k+1}\ad}\Lambda_{\a(k))}~,~k\geq0~.
\eea
In order to generate this equation of motion from a lagrangian we 
must have the term
\bea
\mathcal{L}=\dots +B^{\a(k+2)}\pa_{\a_{k+2}}{}^{\ad}A_{\a(k+1)\ad}+c.c.
\eea
where $B_{\a(k+2)}$ is an extra auxiliary field  with symmetrized indices. The equation 
of motion for $B$ is:
\Ibea{l}
\pa^{\a_{k+2}}{}_{\ad}B_{\a(k+2)}=0  \n\label{B1}\\
~~~~\text{or}\\
\pa^{\a_{k+2}}{}_{\ad}B_{\a(k+2)}\pm c.c.=0~[\text{If we can impose a reality condition}~ A=\pm\bar{A}]\n\label{B2}~.
\Ieea
This equation must also be satisfied identically and give a solution that respects whatever gauge 
transformation $B_{\a(k+2)}$ has. We can repeat the arguments used in $A$ but we quickly realize 
they do not work in this case, because $B$ does not have dotted indices to symmetrize 
over, which was a consequence of the equation of $A$. So equation (\ref{B1}) can not be 
satisfied identically and the only hope for consistence is equation (\ref{B2}). For that to 
work $A$ has to be either real or imaginary, therefore the number of dotted and undotted 
indices of $A$ must match giving $k=0$. It is straight forward to show there is a unique 
solution of (\ref{B2}) and it gives the gauge transformation of $B_{\a\b}$ to be
\Ibea{l}
\delta_g B_{\a\b}=\pa_{(\a}{}^{\bd}L_{\b)\bd}~,~L_{\b\bd}=\mp\bar{L}_{\b\bd}\n
\Ieea
where $L_{\a\ad}$ is not uniquely defined because of the freedom
\Ibea{l}
L_{\b\bd}\sim L_{\b\bd}+\pa_{\b\bd}L~,~L=\mp\bar{L}~~.\n
\Ieea
As can seen from the above construction, there are no fermionic auxiliary fields that in this category.

To conclude, we find that for any massless, $4D,~\mathcal{N}=1$ theory, the supersymmetric
auxiliary fields that appear in its off-shell spectrum organise in the following ways:
\begin{enumerate}
\item[\underline{Fermions}:] There is only one kind of fermionic auxiliary feilds and have the properties\\
~~~(\emph{i}) they come in pairs $(\b,\rho)$~$[\mathcal{L}=\dots+\b^{\dots}\rho_{\dots}+c.c.]$, \\
~~~(\emph{ii}) they are gauge invariant $[\delta_g\b_{\dots}=0=\delta_g\rho_{\dots}]$,\\
~~~(\emph{iii}) they have engineering dimensions $\frac{3}{2}$ and $\frac{5}{2}$.
\item[\underline{Bosons}:] There are three types of auxiliary bosonic fields
\vspace{-1.0ex}
\begin{itemize}
\item[Type (1):] (\emph{i}) they came in pairs of $(A_{\a\ad},B_{\a\b})$ with $A_{\a\ad}=\pm\bar{A}_{\a\ad}$
$~[\mathcal{L}=\dots+B^{\a\b}\pa_{\a}{}^{\bd}A_{\b\bd}+c.c.]$~,\\
(\emph{ii}) they have non-trivial gauge transformations\\
$~~~~~~[~~\delta_g A_{\a\ad}=\pa_{\a\ad}
\lambda~,~\lambda=\pm\bar{\lambda},$\\
$~~~~~~~~~\delta_g B_{\a\b}=\pa_{(\a}{}^{\bd}L_{\b)\bd}~,~\delta_g L_{\b\bd}=\pa_{\b\bd}L~,~L_{\b\bd}
=\mp\bar{L}_{\b\bd}~,~L=\mp\bar{L}~~]$~,\\
(\emph{iii}) they have enginnering dimensions $2$ and $1$.
\item[Type (2):] (\emph{i}) they are singles $[\mathcal{L}=\dots+A^{\dots}A_{\dots}+c.c.]$~,\\
(\emph{ii}) they are gauge invariant $[\delta_g A=0]$~,\\
(\emph{iii}) they have engineering dimensions $2$~.
\item[Type (3):] (\emph{i}) they come in pairs $(A,B)$~$[\mathcal{L}=\dots+A^{\dots}B_{\dots}+c.c.]$~,\\
 (\emph{ii}) they are gauge invariant $[\delta_g A_{\dots}=0=\delta_g B_{\dots}]$~,\\
(\emph{iii}) they have engineering dimensions $3$ and $1$~.
\end{itemize}
\end{enumerate}
It is perhaps illustrative to expand on known examples of the Type (3) behavior already in the 
literature, though these are within the context of lower spin supermultiplets.  The first work to 
demonstrate a pair of bosonic auxiliary fields with unequal engineering dimensions in a 
dynamical action with $\cal N$ = 2 supersymmetry was given in a work completed in 1983 
\cite{RH}.  Since then, a second such $\cal N$ = 2 supersymmetric system \cite{MttR,HypRP} 
has been shown to exist.  As one would expect, the work of \cite{RH} possesses an $\cal N$ 
= 1 truncation and this was done in the work of \cite{D-G} which also gave a generalization 
where an arbitrary number of such pairs were shown to be possible for an $\cal N$ = 1
construction.
%%%%%%%%%%%%%%%%%%%%%%%%%%%%%%%%%
%%%%%%% XXYYXXZZ  XXYYXXZZ   %%%%%%%%%%%%%
\subsection{Matching bosons with fermions}
Every bosonic auxiliary field will be a $(k,l)$-tensor with $k$-undotted and $l$-dotted indices
($k+l$ = even) independently symmetrized. We have also shown that the bosonic auxiliary fields
can be organized into classes $r$ based on their index structure,
their gauge transformation and their reality or not. For each class of tensors we can count the off-shell degrees of 
freedom $\mathcal{D}_{r}$ they carry. Adding all these contributions, 
we can calculate the off-shell degrees of freedom that all auxiliary 
bosons carry
\bea
\mathcal{A}_{B}=\sum_r\mathcal{D}_r\mathcal{N}^{B}_r   \label{Ab}
\eea
The coefficient $\mathcal{N}_{r}$ is a multiplicity factor which counts the number of elements inside each class
and the summation
is over all possible classes. A similar computation can be done for the 
fermionic auxiliary fields and determine $\mathcal{A}_F$. Our \emph{Supersymmetry} requirement to match the 
bosonic and fermionic off-shell d.o.f. can be expressed in the following 
way
\bea
\left(~\parbox{2.5cm}{~integer spin\\ off-shell d.o.f.}~\right)+\mathcal{A}_B=\left(~\parbox{3cm}
{half-integer spin\\ \centering off-shell d.o.f.}~\right)+\mathcal{A}_F~.\label{Dioph}
\eea
This is a Diophantine equation for the coefficients $\mathcal{N}^{B}_r$ and $\mathcal{N}^{F}_r$.
Finding the values of these coefficients will determine completely the off-shell spectrum of the theory.
Of course not all solutions of this Diophantine equation can have this interpretation and we will
not blindly accept all solutions. However, we will show that the rest of our 
requirements will allow us to select only these solutions that correspond to manifestly 
supersymmetric theories. For example, the summation in (\ref{Ab}) takes place over all possible 
types of auxiliary fields. Obviously there are many possibilities and the number of solutions 
that explore all of them becomes very big. Nevertheless, our \emph{Superspace} requirement drastically reduces the allowed 
possibilities. When on top of that we add our demand for \emph{economy}, then we are left with very little options.
These surviving solutions will be the ones that can be realized as supersymmetric theories.

The logic of the arguments above seem ``iron-clad,'' but it is useful to recall one example from
the past demonstrating the possibility of a loop-hole in these arguments.  In the work of \cite{WCG},
it was pointed out that sometimes a symmetry argument combined with arguments about
degrees of freedom, can only be resolved by the essential use of non-linearity.  In this
case, the symmetry was Lorentz symmetry.  However, it does act as a concrete example
where relying solely on the linear realization leads one astray.  With this caveat in mind,
we will continue with our arguments under the assumption that we work in the domain
of purely linear realizations.
%%%%%%%%%%%%%%%%%%%%%%%%%%%%%%%%%%%%%%
%%%%%%%%%%%%%%%%%%%%%%%%%%%%%%%%%%%%%%
%%%%% - Superfield description - %%%%%
\section{Demanding existence of a \emph{Superspace} descriptions}
\subsection{Types of superfields}
The next requirement is that all fields must be generated from a set of superfields, meaning 
there exist an underlying superspace description that makes supersymmetry manifest even 
if we focus only on the component structure of the theory. Surely, this demand must put some 
constraint on the type of tensors that can emerge and therefore affect the balancing of the d.o.f.
through (\ref{Dioph}).

Without loss of generality we can consider all superfields that participate in
the superspace description of the theory to be 
unconstrained. If some of them are not, then we must be able to solve whatever constraint 
they satisfy and express them in terms of unconstrained ones.  This is equivalent to demanding
the existence of a set of so-called ``prepotential superfields'' that provide the most fundamental
description of the theories in questions.  For 4D, $\cal N$ = 1 supersymmetric Yang-Mills
theories, the prepotential is traditionally denoted by $V$, a real unconstrained pseudoscalar
superfield.  All geometrical structures such as connections and field strengths are determined
by this superfield. For 4D, $\cal N$ = 1 supergravity theories, the prepotential is often denoted 
by $U{}^{\g \, \gd}$, a real unconstrained axial vector superfield \cite{OgCur1,OgCur2} leading 
to a geometical formulation \cite{SFSG} of a curved supermanifold.  Supersymmetrical
generalizations of frame fields, spin-connection, and curvature tensor are all fundamentally
defined by the $U{}^{\g \, \gd}$. It was later shown that the prepotential concept applied
to supermultiplets with $Y$ = 1 \cite{MGM2}. 
As these supermultiplets are also gauged ones,
it became obvious that this would be the case for all higher values of $Y$.

Let us consider an arbitrary, unconstrained superfield ${\rm A}_{\a(k)\ad(l)}$ with engineering 
dimensions $d_{{\rm A}}$. The kinetic energy term for this superfield will be quadratic in ${\rm 
A}$ and will have some number $N$ of spinorial covariant derivatives (${\rm D}{}$ and 
${\Bar {\rm D}}$)
\bea
\int d^8z~{\rm A}\underbrace{{\rm D}\dots{\rm D}}_{N}{\rm A}~~\text{or}~\int d^8z~{\rm A}{\rm 
D}\dots{\rm D}\bar{{\rm A}} ~.
\eea
Based on dimensional analysis we must have that
\bea
N=4(1-d_{{\rm A}})~.
\eea
Also, it is easy to show that no matter what the index structure of ${\rm A}$ is, $N$ has to be 
even. Furthermore, $N$ has to be strictly positive, because otherwise ${\rm A}$ will have an 
algebraic equation of motion which means that it can be integrated out and therefore it is irrelevant.
The result is $N
\geq2$, thus $d_{{\rm A}}\leq\tfrac{1}{2}$. The components
\footnote{The various components are labelled by the name of the superfield they come from 
and their position $(n,m)$ in its $\theta$ expansion. For example, $\Phi^{(0,0)}$ is the $\theta$ 
independent term of superfield $\Phi$, $\Phi^{(0,1)}_{\ad}$ is the $\bar{\theta}$ component 
and $\Phi^{(1,1)}_{\a\ad}$ is its $\theta\bar{\theta}$ component. Components with more than 
one index of the same type can be decomposed into symmetric (S) and anti-symmetric (A) 
pieces as $\Phi^{(S)}_{\b\a}=\Phi_{(\b\a)}$~,~$\Phi^{(A)}=C^{\b\a}\Phi_{\b\a}$.}
that such a superfield can generate are:
\begin{center}
\begin{tabular}{l l c l c}
\multicolumn{2}{c}{\centering Group A} & & \multicolumn{2}{c}{\centering Group B}\\ 
\cline{1-2} \cline{4-5}\vspace{-1.5ex}\\
%%%
{}[$d_{{\rm A}}$] & ${\rm A}^{(0,0)}_{\a(k)\ad(l)}$ & & [$d_{{\rm A}}+\tfrac{1}{2}$] & ${\rm 
A}^{(1,0)(S)}_{\a(k+1)\ad(l)}$\vspace{1.5ex}\\
%%%
{}[$d_{{\rm A}}+1$] & ${\rm A}^{(2,0)}_{\a(k)\ad(l)}$ & & [$d_{{\rm A}}+\tfrac{1}{2}$] & ${\rm 
A}^{(1,0)(A)}_{\a(k-1)\ad(l)}$\vspace{1.5ex}\\
%%%
{}[$d_{{\rm A}}+1$] & ${\rm A}^{(0,2)}_{\a(k)\ad(l)}$ & & [$d_{{\rm A}}+\tfrac{1}{2}$] & ${\rm 
A}^{(0,1)(S)}_{\a(k)\ad(l+1)}$\vspace{1.5ex}\\
%%%
{}[$d_{{\rm A}}+1$] & ${\rm A}^{(1,1)(S,S)}_{\a(k+1)\ad(l+1)}$ & & [$d_{{\rm A}}+\tfrac{1}{2}
$] & ${\rm A}^{(0,1)(A)}_{\a(k)\ad(l-1)}$\vspace{1.5ex}\\
%%%
{}[$d_{{\rm A}}+1$] & ${\rm A}^{(1,1)(S,A)}_{\a(k+1)\ad(l-1)}$ & & [$d_{{\rm A}}+\tfrac{3}{2}
$] & ${\rm A}^{(1,2)(S)}_{\a(k+1)\ad(l)}$\vspace{1.5ex}\\
%%%
{}[$d_{{\rm A}}+1$] & ${\rm A}^{(1,1)(A,S)}_{\a(k-1)\ad(l+1)}$ & & [$d_{{\rm A}}+\tfrac{3}{2}
$] & ${\rm A}^{(1,2)(A)}_{\a(k-1)\ad(l)}$\vspace{1.5ex}\\
%%%
{}[$d_{{\rm A}}+1$] & ${\rm A}^{(1,1)(A,A)}_{\a(k-1)\ad(l-1)}$ & & [$d_{{\rm A}}+\tfrac{3}{2}
$] & ${\rm A}^{(2,1)(S)}_{\a(k)\ad(l+1)}$\vspace{1.5ex}\\
%%%
{}[$d_{{\rm A}}+2$] & ${\rm A}^{(2,2)}_{\a(k)\ad(l)}$ & & [$d_{{\rm A}}+\tfrac{3}{2}$] & ${\rm 
A}^{(2,1)(A)}_{\a(k)\ad(l-1)}$\vspace{2ex}\\
\multicolumn{5}{c}{Table 1: List of fields generated by superfield ${\rm A}$}
%%%
\end{tabular}
\end{center}
Groups A and B are the groups of bosonic or the fermionic components depending if superfield ${\rm A}$ is bosonic 
($k+l$ = even) or fermionic ($k+l$ = odd). All elements of group A have the same grassman
parity, equal to that of the superfield A and in contrast the elements of group B
have the opposity grassman parity. If 
${\rm A}$ is complex it carries $16kl+16k+16l+16$ bosons and equal fermions, whereas if it is real (that also means $k=l$)
it will carry $8k^2+16k+8$ bosons and equal fermions.
An obvious observation is that the range of engineering dimensions for the components is from 
$d_{{\rm A}}$ to $d_{{\rm A}}+2$. Therefore in order for a superfield to provide propagating bosons and fermions we must 
have $d_{{\rm A}}+2\geq\tfrac{3}{2}\Rightarrow~d_{{\rm A}}\geq-\tfrac{1}{2}$. Therefore we 
conclude that the allowed superfields must have dimensions $d$ with
\bea
-\tfrac{1}{2}\leq d\leq\tfrac{1}{2}~.
\eea
So all bosonic 
superfields will have $d=0$ and the fermionic superfields will have $d=\tfrac{1}{2}$ or $d=-
\tfrac{1}{2}$. The conclusion is:
\begin{enumerate}
\item[$\alpha$)] For ${\cal N}=1$ theories, bosonic auxiliary fields of type ($3$) are not permitted.
%\\
%Therefore auxiliary bosons must be either
%type (1) [pairs of a real $(1,1)$-tensor with a $(2,0)
%$-tensor with gauge transformations and dimension 2 and 1 respectively] or \\
%type (2) [gauge 
%invariant tensors of any rank with dimension 2]~,
\item[$\beta$)] Fermionic auxiliary fields can only be generated by fermionic superfields with enginnering dimensions $d=
\tfrac{1}{2}$~.
\end{enumerate}
%%%%%%%%%%%%%%%%%%%%%%%%%%%%%%%%%
%%%%%%% XXYYXXZZ  XXYYXXZZ   %%%%%%%%%%%%%
\subsection{Gauge transformations}
It is a fact of physics that there is a discontinuity in the degrees of freedom of massless and 
massive theories and that insisting on describing the system in a local manner forces upon 
us the concept of redundancies (gauge symmetries). Hence, for massless theories there must be a gauge symmetry that can be lifted all the way to the superspace description. 
Here we discuss the various options for the superspace gauge symmetry of an arbitrary superfield 
${\rm A}_{\a(k)\ad(l)}$.

First of all, the transformation of a superfield must not include algebraic terms bacause then 
we have the freedom to completely remove it, thus making it irrelevant.
Therefore, there must be at least one spinorial covariant derivative and the most general gauge 
transformation of ${\rm A}_{\a(k)\ad(l)}$ can be parametrized in the following way\footnote{One 
must keep in mind that for the special case of $k=0$ or $l=0$ one must consider the presence of $\D^2$
or $\Dd^2$ terms in the transformation law}:
\bea
\delta{\rm A}_{\a(k)\ad(l)}=&~~{\rm D}_{(\a_k}L_{\a(k-1))\ad(l)}+{\Bar {\rm D}}_{(\ad_{l}}\Lambda_{
\a(k)\ad(l-1))}\\
&+{\rm D}^{\b}J_{\b\a(k)\ad(l)}+{\Bar {\rm D}}^{\bd}I_{\a(k)\bd\ad(l)}
\eea
for some $L_{\a(k-1)\ad(l)},~\Lambda_{\a(k)\ad(l-1)},~J_{\a(k+1)\ad(l)},~I_{\a(k)\ad(l+1)}$. The 
first two terms consist what we will call a type $\mathcal{I}$ transformation and the last two terms 
is a type $\mathcal{II}$ transformation and any other type of gauge transformation can be generated 
by combining them and selecting appropriately $L,~\Lambda,~J,~I$. Now we can deduce the 
gauge transformation of the various components of ${\rm A}$:
\begin{center}
\vspace{-3ex}
\resizebox{\textwidth}{!}{\small
\renewcommand{\arraystretch}{1.9}
\begin{tabular}{l || c | c | c | c}
\multicolumn{1}{l}{Table 2:} & \multicolumn{4}{c}{\underline{~Group A~}}\vspace{-2ex}\\
\multicolumn{1}{c}{{}} & \multicolumn{1}{c}{${\rm D}_{(\a_k}L_{\a(k-1))\ad(l)}$} & \multicolumn{
1}{c}{${\Bar {\rm D}}_{(\ad_{l}}\Lambda_{\a(k)\ad(l-1))}$} & \multicolumn{1}{c}{${\rm D}^{\b}J_{
\b\a(k)\ad(l)}$} & \multicolumn{1}{c}{${\Bar {\rm D}}^{\bd}I_{\a(k)\bd\ad(l)}$}\\
\hline
${\rm A}^{(0,0)}_{\a(k)\ad(l)}$  & ${\rm D}_{(\a_k}L_{\a(k-1))\ad(l)}|$ & ${\Bar {\rm D}}_{(\ad_{l
}}\Lambda_{\a(k)\ad(l-1))}|$ & ${\rm D}^{\b}J_{\b\a(k)\ad(l)}|$ & ${\Bar {\rm D}}^{\bd}I_{\a(k)\bd
\ad(l)}|$  \\ \hline
${\rm A}^{(2,0)}_{\a(k)\ad(l)}$  & 0  & ${\rm D}^2{\Bar {\rm D}}_{(\ad_{l}}\Lambda_{\a(k)\ad(l-1
))}|$ & 0  & ${\rm D}^2{\Bar {\rm D}}^{\bd}I_{\a(k)\bd\ad(l)}|$ \\ \hline
 ${\rm A}^{(0,2)}_{\a(k)\ad(l)}$  & ${\Bar {\rm D}}^2{\rm D}_{(\a_k}L_{\a(k-1))\ad(l)}|$ & 0 & ${\Bar 
 {\rm D}}^2{\rm D}^{\b}J_{\b\a(k)\ad(l)}|$ & 0  \\ \hline
${\rm A}^{(1,1)(S,S)}_{\a(k+1)\ad(l+1)}$  & $i\pa_{(\a_{k+1}(\ad_{l+1}}{\rm D}_{\a_k}L_{\a(k-1
))\ad(l))}|$ & $i\pa_{(\a_{k+1}(\ad_{l+1}}{\Bar {\rm D}}_{\ad_{l}}\Lambda_{\a(k))\ad(l-1))}|$ & 
\parbox{4.7cm}{\centering $i\pa_{(\a_{k+1}(\ad_{l+1}}{\rm D}^{\b}J_{\b\a(k))\ad(l))}|$\\ ${\Bar 
{\rm D}}_{(\ad_{l+1}}{\rm D}^2J_{\a(k+1)\ad(l))}|$} & \parbox{4.7cm}{\centering $i\pa_{(\a_{k
+1}(\ad_{l+1}}{\Bar {\rm D}}^{\bd}I_{\a(k))\bd\ad(l))}|$\\ ${\rm D}_{(\a_{k+1}}{\Bar {\rm D}}^2I_{
\a(k))\ad(l+1)}|$} \\ \hline
${\rm A}^{(1,1)(S,A)}_{\a(k+1)\ad(l-1)}$  & $i\pa_{(\a_{k+1}}{}^{\ad_{l}}{\rm D}_{\a_k}L_{\a(k-1
))\ad(l)}|$ & \parbox{4.7cm}{\centering $i\pa_{(\a_{k+1}}{}^{\ad_l}{\Bar {\rm D}}_{(\ad_{l}}
\Lambda_{\a(k))\ad(l-1))}|$\\ ${\rm D}_{(\a_{k+1}}{\Bar {\rm D}}^2\Lambda_{\a(k))\ad(l-1)}|$
} & \parbox{4.7cm}{\centering $i\pa_{(\a_{k+1}}{}^{\ad_l}{\rm D}^{\b}J_{\b\a(k))\ad(l)}|$\\ ${
\Bar {\rm D}}^{\ad_l}{\rm D}^2J_{\a(k+1)\ad(l)}|$} & $i\pa_{(\a_{k+1}}{}^{\gd}{\Bar {\rm D}}^{
\bd}I_{\a(k))\gd\bd\ad(l-1)}|$ \\ \hline
${\rm A}^{(1,1)(A,S)}_{\a(k-1)\ad(l+1)}$  & \parbox{4.7cm}{\centering $i\pa^{\a_k}{}_{(\ad_{
l+1}}{\rm D}_{(\a_k}L_{\a(k-1))\ad(l))}|$\\ ${\Bar {\rm D}}_{(\ad_{l+1}}{\rm D}^2L_{\a(k-1)\ad(l))
}|$} & $i\pa^{\a_k}{}_{(\ad_{l+1}}{\Bar {\rm D}}_{\ad_{l}}\Lambda_{\a(k)\ad(l-1))}|$ & $i\pa^{\g}
{}_{(\ad_{l+1}}{\rm D}^{\b}J_{\g\b\a(k-1)\ad(l))}|$ & \parbox{4.7cm}{\centering $i\pa^{\g}{}_{(
\ad_{l+1}}{\Bar {\rm D}}^{\bd}I_{\g\a(k-1))\bd\ad(l))}|$\\ ${\rm D}^{\g}{\Bar {\rm D}}^2I_{\g\a(k-1
)\ad(l+1)}|$} \\ \hline
${\rm A}^{(1,1)(A,A)}_{\a(k-1)\ad(l-1)}$  & \parbox{4.7cm}{\centering $i\pa^{\a_k\ad_l}{\rm 
D}_{(\a_k}L_{\a(k-1))\ad(l)}|$\\ ${\Bar {\rm D}}^{\ad_l}{\rm D}^2L_{\a(k-1)\ad(l)}|$} & \parbox{
4.7cm}{\centering $i\pa^{\a_k\ad_l}{\Bar {\rm D}}_{(\ad_{l}}\Lambda_{\a(k)\ad(l-1))}|$\\ ${\rm 
D}^{\a_k}{\Bar {\rm D}}^2\Lambda_{\a(k)\ad(l-1)}|$} & $i\pa^{\g\gd}{\rm D}^{\b}J_{\g\b\a(k-1)
\gd\ad(l-1)}|$ & $i\pa^{\g\gd}{\Bar {\rm D}}^{\bd}I_{\g\a(k-1)\gd\bd\ad(l-1)}|$ \\ \hline
${\rm A}^{(2,2)}_{\a(k)\ad(l)}$  & \parbox{4.9cm}{\centering $i\pa^{\b\bd}[{\rm D}_{\b},{\Bar {
\rm D}}_{\bd}]{\rm D}_{(\a_k}L_{\a(k-1))\ad(l)}|$\\ $\Box{\rm D}_{(\a_{k}}L_{\a(k-1)\ad(l)}|$} & 
\parbox{4.9cm}{\centering $i\pa^{\b\bd}[{\rm D}_{\b},{\Bar {\rm D}}_{\bd}]{\Bar {\rm D}}_{(\ad_l
}\Lambda_{\a(k)\ad(l-1))}|$\\ $\Box{\Bar {\rm D}}_{(\ad_{l}}\Lambda_{\a(k)\ad(l-1))}|$} & \parbox{
4.7cm}{\centering $i\pa^{\g\gd}[{\rm D}_{\g},{\Bar {\rm D}}_{\gd}]{\rm D}^{\b}J_{\b\a(k)\ad(l)}|$\\ 
$\Box{\rm D}^{\b}J_{\b\a(k)\ad(l)}|$} & \parbox{4.7cm}{\centering $i\pa^{\g\gd}[{\rm D}_{\g},{
\Bar {\rm D}}_{\gd}]{\Bar {\rm D}}^{\bd}I_{\a(k))\bd\ad(l))}|$\\ $\Box{\Bar {\rm D}}^{\bd}I_{\a(k)
\bd\ad(l)}|$} 
\end{tabular}
}
\end{center}
%%%
%%%
\begin{center}
\resizebox{\textwidth}{!}{\small
\renewcommand{\arraystretch}{1.9}
\begin{tabular}{l || c | c | c | c}
\multicolumn{1}{l}{Table 3:} & \multicolumn{4}{c}{\underline{~Group B~}}\vspace{-2ex}\\
\multicolumn{1}{c}{{}} & \multicolumn{1}{c}{${\rm D}_{(\a_k}L_{\a(k-1))\ad(l)}$} & 
\multicolumn{1}{c}{${\Bar {\rm D}}_{(\ad_{l}}\Lambda_{\a(k)\ad(l-1))}$} & \multicolumn{1}{c}{
${\rm D}^{\b}J_{\b\a(k)\ad(l)}$} & \multicolumn{1}{c}{${\Bar {\rm D}}^{\bd}I_{\a(k)\bd\ad(l)}$}\\
\hline
${\rm A}^{(1,0)(S)}_{\a(k+1)\ad(l)}$ & 0 & ${\rm D}_{(\a_{k+1}}{\Bar {\rm D}}_{(\ad_l}
\Lambda_{\a(k))\ad(l-1))}|$ & ${\rm D}^2J_{\a(k+1)\ad(l)}|$ & ${\rm D}_{(\a_{k+1}}{\Bar 
{\rm D}}^{\bd}I_{\a(k))\bd\ad(l)}|$ \\ \hline
%%%
${\rm A}^{(1,0)(A)}_{\a(k-1)\ad(l)}$ & ${\rm D}^2L_{\a(k-1)\ad(l)}|$ & ${\rm D}^{\b}{\Bar 
{\rm D}}_{(\ad_l}\Lambda_{\b\a(k-1)\ad(l-1))}|$ & 0 & ${\rm D}^{\b}{\Bar {\rm D}}^{\bd}
I_{\b\a(k-1)\bd\ad(l)}|$ \\ \hline
%%%
${\rm A}^{(0,1)(S)}_{\a(k)\ad(l+1)}$ & ${\Bar {\rm D}}_{(\ad_{l+1}}{\rm D}_{(\a_k}L_{\a(k-1))
\ad(l))}|$ & 0 & ${\Bar {\rm D}}_{(\ad_{l+1}}{\rm D}^{\b}J_{\b\a(k)\ad(l))}|$ & ${\Bar {\rm D}}^2
I_{\a(k)\ad(l+1)}|$ \\ \hline
%%%
${\rm A}^{(0,1)(A)}_{\a(k)\ad(l-1)}$ & ${\Bar {\rm D}}^{\bd}{\rm D}_{(\a_k}L_{\a(k-1))\bd\ad(
l-1)}|$ & ${\Bar {\rm D}}^2\Lambda_{\a(k)\ad(l-1)}|$ & ${\Bar {\rm D}}^{\bd}{\rm D}^{\b}J_{
\b\a(k)\bd\ad(l-1)}|$ & 0 \\ \hline
%%%
${\rm A}^{(1,2)(S)}_{\a(k+1)\ad(l)}$ & $i\pa_{(\a_{k+1}}{}^{\bd}{\Bar {\rm D}}_{\bd}{\rm D}_{
\a_{k}}L_{\a(k-1))\ad(l)}$ & $i\pa_{(\a_{k+1}(\ad_{l}}{\Bar {\rm D}}^2\Lambda_{\a(k))\ad(l-1))
}|$ & \parbox{4.7cm}{\centering $i\pa_{(\a_{k+1}}{}^{\bd}{\Bar {\rm D}}_{\bd}{\rm D}^{\b}J_{
\b\a(k))\ad(l)}|$ \\ ${\Bar {\rm D}}^2{\rm D}^2J_{\a(k+1)\ad(l)}|$} & $i\pa_{(\a_{k+1}}{}^{\bd}{
\Bar {\rm D}}^2 I_{\a(k))\bd\ad(l)}|$ \\ \hline
%%%
${\rm A}^{(1,2)(A)}_{\a(k-1)\ad(l)}$ & \parbox{4.7cm}{\centering $i\pa^{\a_k\bd}{\Bar {\rm D}}_{
\bd}{\rm D}_{(\a_k}L_{\a(k-1))\ad(l)}$ \\ ${\Bar {\rm D}}^2{\rm D}^2L_{\a(k-1)\ad(l)}|$} & $i\pa^{\b
}{}_{(\ad_l}{\Bar {\rm D}}^2\Lambda_{\b\a(k-1)\ad(l-1))}|$ & $i\pa^{\b\bd}{\Bar {\rm D}}_{\bd}{\rm 
D}^{\g}J_{\g\b\a(k-1)\ad(l)}$ & $i\pa^{\b\bd}{\Bar {\rm D}}^2I_{\b\a(k-1)\bd\ad(l)}|$ \\ \hline
%%%
${\rm A}^{(2,1)(S)}_{\a(k)\ad(l+1)}$ & $i\pa_{(\a_k (\ad_{l+1}}{\rm D}^2L_{\a(k-1))\ad(l))}|$ & $i
\pa^{\b}{}_{(\ad_{l+1}}{\rm D}_{\b}{\Bar {\rm D}}_{\ad_l}\Lambda_{\a(k)\ad(l-1))}|$ & $i\pa^{\b}{
}_{(\ad_{l+1}}{\rm D}^2 J_{\b\a(k)\ad(l))}|$ & \parbox{4.7cm}{\centering $i\pa^{\b}{}_{(\ad_{l+1}}
{\rm D}_{\b}{\Bar {\rm D}}^{\bd}I_{\a(k)\bd\ad(l))}|$\\ ${\rm D}^2{\Bar {\rm D}}^2I_{\a(k)\ad(l+1)}|$
} \\ \hline
%%%
${\rm A}^{(2,1)(A)}_{\a(k)\ad(l-1)}$ & $i\pa_{(\a_k}{}^{\bd}{\rm D}^2L_{\a(k-1))\bd\ad(l-1)}|$ & 
\parbox{4.7cm}{\centering $i\pa^{\b\ad_l}{\rm D}_{\b}{\Bar {\rm D}}_{(\ad_l}\Lambda_{\a(k)
\ad(l-1)}|$\\ ${\rm D}^2{\Bar {\rm D}}^2\Lambda_{\a(k)\ad(l-1)}|$} & $i\pa^{\b\bd}{\rm D}^2 
J_{\b\a(k)\bd\ad(l-1)}|$ & $i\pa^{\b\bd}{\rm D}_{\b}{\Bar {\rm D}}^{\gd}I_{\b\a(k-1)\bd\gd\ad(
l-1)}|$ \\ \hline
\end{tabular}
}
\end{center}

The above tables provide a list of the terms that appear in the transformation law of each of 
the components of ${\rm A}$, for every possible type of gauge transformation. Notice that there
are two types of terms. The terms that have spacetime derivatives and all the rest, which we will call 
algebraic because they include algebraic terms of components of the gauge parameters\footnote{
One must be careful because the gauge parameters may have some further ${\rm D}$-structure 
inside them that will convert the seemingly algebraic terms to derivative terms or make them 
vanish.}. The algebraic terms could be used to gauge remove the corresponding ${\rm A}$ 
component. In other words, the components that have only derivative terms in their transformation 
can not be removed. Such a term is the component ${\rm A}^{(2,2)}_{\a(k)\ad(l)}$. For any type of 
transformation, the coefficient of $\theta^2\bar{\theta}^2$ can never be gauged away and that is 
obvious because it has the maximum amount of $\theta$s and $\bar{\theta}$s hence it can not include algebraic terms
in its transformation since the superfield transformation law is not algebraic.
Therefore for bosonic superfields ($d=0$) this component will provide 
an auxiliary boson and for fermionic superfields with $d=\tfrac{1}{2}$ the ${\rm A}^{(2,2)}$ component 
gives the one member ($\b$) of the pair of auxiliary fermions.

The usefulness of these two tables is twofold.
First of all, they introduce various associations among the components of a superfield. These associations, provide
a set a conditional constraints that will help us reduce the number of possible solutions of the corresponding Diophantine equation.
Specifically, we can deduce that if some component of a superfield can be gauged removed then there is a set of other components that
can be removed as well and therefore a potential solution of the Diophantine equation that does not comply with these conditions can not
have the interpretation of a manifest supersymmetric theory and will not be accepted. To generate the various sets of conditional
constraints we focus at the components ${\rm A}^{(1,2)(S)}_{\a(k+1)\ad(l)},~{\rm A}^{(1,2)(A)}_{\a(k-1)\ad(l)},~{\rm A}^{(2,1)(S)}_{\a(k)\ad(l+1)},~{\rm A}^{(2,1)(A)}_{\a(k)\ad(l-1)}$. These component are special, because for each one of them there is a unique way to gauge remove them partially or fully and therefore they demand the presence of a specific term in the transformation law of the superfield. Hence we can safely conclude that there will be a set of other components that can be (partially) eliminated by the same term. The results are:
\begin{center}
\vspace{-5ex}
\resizebox{\textwidth}{!}{\small
\renewcommand{\arraystretch}{1.9}
\begin{tabular}{| c | c |}
\multicolumn{2}{l}{Table4:~~\underline{Conditional Constraints}}\vspace{-5ex}\\
\multicolumn{2}{c}{{}}\\
\hline
 If this is (partially) removed & Then these are (partially) removed as well\\
\hline
${\rm A}^{(1,2)(S)}_{\a(k+1)\ad(l)}$ & ${\rm A}^{(1,0)(S)}_{\a(k+1)\ad(l)}$,~${\rm A}^{(0,1)(S)}_{\a(k)\ad(l+1)}$,~${\rm A}^{(0,1)(A)}_{\a(k)\ad(l-1)}$,~
${\rm A}^{(0,0)}_{\a(k)\ad(l)}$,~${\rm A}^{(0,2)}_{\a(k)\ad(l)}$,~${\rm A}^{(1,1)(S,S)}_{\a(k+1)\ad(l+1)}$,~${\rm A}^{(1,1)(S,A)}_{\a(k+1)\ad(l-1)}$\\
\hline
%%%%%%%%%%%%%%%%%%%%%%%%%%%%%%%%
%%%%%%%%%%%%%%%%%%%%%%%%%%%%%%%%
${\rm A}^{(1,2)(A)}_{\a(k-1)\ad(l)}$ & ${\rm A}^{(1,0)(A)}_{\a(k-1)\ad(l)}$,~${\rm A}^{(0,1)(S)}_{\a(k)\ad(l+1)}$,~${\rm A}^{(0,1)(A)}_{\a(k)\ad(l-1)}$,~
${\rm A}^{(0,0)}_{\a(k)\ad(l)}$,~${\rm A}^{(0,2)}_{\a(k)\ad(l)}$,~${\rm A}^{(1,1)(A,S)}_{\a(k-1)\ad(l+1)}$,~${\rm A}^{(1,1)(A,A)}_{\a(k-1)\ad(l-1)}$\\
\hline
%%%%%%%%%%%%%%%%%%%%%%%%%%%%%%%%
%%%%%%%%%%%%%%%%%%%%%%%%%%%%%%%%
${\rm A}^{(2,1)(S)}_{\a(k)\ad(l+1)}$ & ${\rm A}^{(1,0)(S)}_{\a(k+1)\ad(l)}$,~${\rm A}^{(1,0)(A)}_{\a(k-1)\ad(l)}$,~${\rm A}^{(0,1)(S)}_{\a(k)\ad(l+1)}$,~
${\rm A}^{(0,0)}_{\a(k)\ad(l)}$,~${\rm A}^{(2,0)}_{\a(k)\ad(l)}$,~${\rm A}^{(1,1)(S,S)}_{\a(k+1)\ad(l+1)}$,~${\rm A}^{(1,1)(A,S)}_{\a(k-1)\ad(l+1)}$\\
\hline
%%%%%%%%%%%%%%%%%%%%%%%%%%%%%%%%
%%%%%%%%%%%%%%%%%%%%%%%%%%%%%%%%
${\rm A}^{(2,1)(A)}_{\a(k)\ad(l-1)}$ & ${\rm A}^{(1,0)(S)}_{\a(k+1)\ad(l)}$,~${\rm A}^{(1,0)(A)}_{\a(k-1)\ad(l)}$,~${\rm A}^{(0,1)(A)}_{\a(k)\ad(l-1)}$,~
${\rm A}^{(0,0)}_{\a(k)\ad(l)}$,~${\rm A}^{(2,0)}_{\a(k)\ad(l)}$,~${\rm A}^{(1,1)(S,A)}_{\a(k+1)\ad(l-1)}$,~${\rm A}^{(1,1)(A,A)}_{\a(k-1)\ad(l-1)}$\\
\hline
\end{tabular}
}
\end{center}

The second use of tables 2 and 3 is to help us determine the gauge transformations we need to have given the set of components we find by solving the Diophantine equation.
%%%%%%%%%%%%%%%%%%%%%%%%%%%%%%%%%%%%%%
%%%%%%%%%%%%%%%%%%%%%%%%%%%%%%%%%%%%%%
%%%%% - Highest spin economy - %%%%%%%
\section{Demanding \emph{Economy}}
For theories without supersymmetry ($\mathcal{N}=0$) given a tensor of some rank, 
it is natural to attempt to use it for the description of the highest
possible spin it contains. From this point of view, the various on-shell constraints 
it must satisfy and the gauge transformation it has  (for massless theories) it is just a way to remove the
lower spin irreps it also contains. This is the preferred choice
to describe any spin, although one could use an even higher rank tensor to describe lower spins, (e.g. \cite{HSLS} for an explicit discussion for such an approach). This approach involves a sense of economy, meaning that we introduce the minimal set of d.o.f required to describe a specific spin with all the symmetries manifest. Also it is one of the principles that lead to the off-shell spectrum of massless higher spins discussed in section 2. This economical description is now elevated to the ${\cal N}=1$ case and becomes
one of our requirements. Therefore, we will search for the most economical solutions of the Diophantine equation in terms of the number of superfields we require to have, their rank and the degrees of freedom they carry off-shell. 
The consequences of having an economical description of a higher spin supersymmetric theory
are:
\begin{enumerate}
\item[\emph{i})] the component that describes the highest spin must be unique and
\item[\emph{ii})] no other component tensor has higher rank 
\end{enumerate}
In Table 1 notice that the component ${\rm A}^{(1,1)(S,S)}_{\a(k+1)\ad(l+1)}$ is the highest rank 
component present
and it is unique. These characteristics fit exactly to our demands, therefore for an economical description 
it makes sense to identify this component with the highest 
spin component of the on-shell spectrum. The consequence of this identification is to fix the values for the integers
$k,~l,~d$ and the reality or not of 
the superfield that will carry the highest spin. 

For example, if the highest spin is integer $s$ then we must have
\Ibea{l}
{\rm A}^{(1,1)(S,S)}_{\a(k+1)\ad(l+1)}\sim h_{\a(s)\ad(s)}\vspace{1ex}\\
\hspace{8ex}k+1=s\\
\hspace{8ex}l+1=s\n\\
\hspace{8ex}d+1=1
\Ieea
thus the superfield that will carry it\footnote{The superfield that carries the 
highest spin, we will call it the \emph{main} one.} must be a real bosonic superfield $H_{\a(s-1)\ad(s-1)}$ 
with $d_H=0$. Based on the results of the 
previous section we can immediatly say that one of the auxiliary bosons must be a real $(s-1,
s-1)$-tensor which corresponds to the $H^{(2,2)}_{\a(s-1)\ad(s-1)}$. Also we can conclude that
if any other superfield is required for the description of the theory must have less rank.

Similarly for the case of half-integer $s+\tfrac{1}{2}$ highest spin we must do the following identification
\Ibea{l}
{\rm A}^{(1,1)(S,S)}_{\a(k+1)\ad(l+1)}\sim \psi_{\a(s+1)\ad(s)}\vspace{1ex}\\
\hspace{8ex}k+1=s+1\\
\hspace{8ex}l+1=s\n\\
\hspace{8ex}d+1=\tfrac{3}{2}
\Ieea
The answer is that the main 
superfield must be a fermionic $\Psi_{\a(s)\ad(s-1)}$ with $d_{\Psi}=\tfrac{1}{2}$. Also, we conclude that there must be at 
least two auxiliary fermionic $(s, \, s-1)$-tensors which are gauge invariant, one of which corresponds to $\Psi^{(2,2)}_{\a(s)\ad(s-1)}$.

Furthermore, the highest spin component must have a specific gauge transformation\footnote{See 
section 2.}. From Table 2 we conclude that in order to match the expected higher spin transformation 
law, without having the risk to be able to gauge remove ${\rm A}^{(1,1)(S,S)}_{\a(k+1)\ad(l+1)}$, the \emph{main} 
superfield that carries the highest spin component must have a transformation of type $\mathcal{I}$.

To gain a deeper appreciation of the restrictive nature of the use of the economical assumption,
it is useful to recall the results in \cite{K2}, where some of the results violate this assumption.
To this point in time, this behaviour is only known to happen in a restricted class of the $(s, \, s+1/2)$ supermultiplets and
is filtered out by the assumption of economic higher spin description.
%%%%%%%%%%%%%%%%%%%%%%%%%%%%%%%%%%%%%%
%%%%%%%%%%%%%%%%%%%%%%%%%%%%%%%%%%%%%%
%%%%%%%% - Vector multiplet - %%%%%%%%
\section{$Y=1/2$ - Vector supermultiplet}
In the previous three sections we presented the basic requirements we want to impose and derived their consequences. 
The next step is to apply them to various supermultiplets, starting with the vector supermultiplet, which has superspin $Y=\tfrac{1}{2}$.
Now we go through the 
following steps:\vspace{1ex}\\
1 - The dynamics involve a spin $1$ and a spin $1/2$. The off-shell d.o.f. provided by the 
spins are
\Ibea{l}
\text{spin}~1 : (s^2+2)|_{s=1} = 3~,\\
\text{spin}~1/2 : (4s^2+4s+4)|_{s=0} = 4~.
\Ieea
\vspace{-3ex}\\
2 - The highest spin is $1$, hence the 
main superfield in the superspace description must be a real scalar superfield $H$ with 
dimensions $0$, which also provides one real scalar auxiliary boson $(H^{(2,2)})$. Because 
it is scalar it can not be of type (1) hence it must be of type (2) and carries $1$ off-shell d.o.f.
$~\Rightarrow$ ~$\mathcal{A}_{B}=1$ and $\mathcal{A}_{F}=0$.
\vspace{1ex}\\
3 - We check for the matching of bosons with fermions
\bea
3+1 = 4+0
\eea
and since they do match, we do not have to add any more auxiliary components. So the theory will be 
described only by the main superfield $H$.\vspace{1ex}\\
4 - The gauge transformation of $H$ is determined by the fact that the component $H^{(0,0)},
~H^{(2,0)},~H^{(1,0)}_{\a}$ must be gauge removed, component $H^{(2,2)}$ must be gauge 
invariant and the reality condition of $H$. The unique answer is
\bea
\delta_g H = {\rm D}{}^2\bar{L}+{\Bar {\rm D}}^2L~.
\eea
%
%
%This simple example also indicates the future need to extend our discussions to
%include ``variant representations.'' In an old work \cite{VarSF}, it was shown the existence
%of a much less recognized 4D, $\cal N$ = 1 supermultiplet that possesses the component
%field content: $V_{\a\ad}$, $\chi_{\a}$, ${\cal 
%G}_{\a\ad}$ that possesses the gauge symmetries
%\bea 
%\delta_g  V_{\a\ad} ~=~ \partial_{\a\ad} \lambda ~~~~,~~~~
%\delta_g  {\cal G}_{\a\ad}~=~ \pa{}_{\a\ad} {\cal K}^{\ad\, \bd} ~+~ {\rm {h. \, c.~~~~with}} ~~~~~~
% {\cal K}^{\ad\, \bd} ~=~  {\cal K}^{\bd\, \ad} ~~,
% \label{3F}
%\eea
%so that the component field $V_{\a\ad} $ is indeed a gauge field for a U(1) symmetry
%and component field ${\cal G}_{\a\ad} $ is the Hodge-dual formulation of a gauge
%3-form field.  The ``main'' superfield that describes this supermultiplet is a spinor
%$\Xi{}_{\a}$ which satisfies the constraint ${{\rm D}}{}_{\a}$  $\Xi{}_{\b}$ = 0.  The
%superfield formulation of the gauge transformations in (\ref{3F}) takes the form
%\bea
%\delta_g \Xi{}_{\a}  ~=~ {{\rm D}}{}_{\a}  \left(  \,  {{\rm D}}{}^{\a} \, \L_{\a} ~+~
% {\Bar {\rm D}}{}^{\ad} \, {\Bar \L}_{\ad} 
%\, \right)  ~~~ 
% {\rm {with}} ~~~~   {{\rm D}}{}_{\a} {\Bar \L}_{\ad} ~=~ 0 ~~~.
%\eea
%
%The fact that straightforward application of our algorithm is mute on the existence
%of this supermultiplets implies the need for its additional development.
%%%%%%%%%%%%%%%%%%%%%%%%%%%%%%%%%%%%%%
%%%%%%%%%%%%%%%%%%%%%%%%%%%%%%%%%%%%%%
%%%%%%%% - Gravitino-matter multiplet - %%%%%%%%
\section{$Y=1$ - Matter-Gravitino supermultiplets}
Next is the matter-gravitino supermultiplets with $Y=1$.
In this case we have:\\
1 - The dynamic part of the theory is that of free spins $3/2$ and $1$, which provide the following number
of off-shell d.o.f.
\Ibea{l}
\text{spin}~1 : (s^2+2)|_{s=1} = 3~,\n\label{spin1}\\
\text{spin}~3/2 : (4s^2+4s+4)|_{s=1} = 12~.\n
\Ieea
\vspace{-3ex}\\
2 - The highest spin is $3/2$, therefore the main superfield must be a $(1,0)$-tensor $\Psi_{\a}$
of engineering dimensions $1/2$ and it will provide a fermionic auxiliary component $\Psi_{\a}^{(2,2)}$ with dimensions $\tfrac{5}{2}$ and since
we know that fermionic auxiliary fields come in pairs, there must be another fermionic auxiliary component with the same index structure and dimensions $\tfrac{3}{2}$.
In detail, superfield $\Psi_{\a}$ has the following list of potential auxiliary components:
\begin{center}
\begin{tabular}{l r l c c c}
 $1~\times~\Psi_{\a}~,~[\Psi_{\a}]=\tfrac{1}{2}$:~& & Tensor & Dimensions & off-shell d.o.f. & multiplicity\\
& Fermions: & $(1,0)$ & [$\tfrac{5}{2}$] & $4$ & 1 \\
 &        & $(1,0)$ & [$\tfrac{3}{2}$] & $4$ & $\le3$ \\
{}\\
 & Bosons: & $(2,0)$ & [$2$] & $6$ [type (2)] & $\le1$ \\
 &  &  & [$2$] & $3$ [type (1)] & $\le1$\\
 &  &  & [$1$] & $3$ [type (1)] & $\le1$ \\
 &  & $(0,0)$ & [$2$] & $1$ [type (2)] & $\le2$\\
 &  & $(1,1)$ & [$2$] & $4$ [type (2)] & $\le2$ \\
 &  &  & [$2$] & $3$ [type (1)] & $\le2$ \\
 &  &  & [$1$] & $3$ [type (1)] & $\le2$
\end{tabular}
\end{center}
where the last column gives the number of times a specific type of a component appears.
If the theory requires extra fields, respecting the uniqueness of higher spin, then these extra superfields must be lower rank tensors.
In this case there is no alternative but for them to be ${\cal N}_V$ copies of a real scalar. The potential auxiliary
components they can provide are:
\begin{center}
\begin{tabular}{l r l c c c}
${\cal N}_V~\times~V~,~[V]=0$:~& & Tensor & Dimensions & off-shell d.o.f. & multiplicity \\
& Bosons: & $(0,0)$ & [$2$] & $1$ [type (2)] & ${\cal N}_V$\\
& & $(1,1)$ & [$1$] & $3$ [type (1)] & $\le {\cal N}_V$
\end{tabular}
\end{center}
3 - Matching bosons with fermions:\\
For fermions we can immediately write that $\mathcal{A}_{F}=8$, since there is a non-removable $(1,0)$ component coming from $\Psi_{\a}$ and we know that auxiliary fermions must appear in pairs. Therefore, the total number of off-shell d.o.f for this supermultiplet will be $12+8=20$ and the rest of potential auxiliary fermionic components must be removed by an appropriate gauge symmetry. For the bosons we have to consider the two different types we can have [type (1) and type (2)]. The potential type (2) auxiliary fields are:
\Ibea{l}
\mathcal{N}_{(2,0)}={\cal N}_{(2,0)}^{\Psi}~,~0\leq{\cal N}_{(2,0)}^{\Psi}\leq 1~,\n   \\
\mathcal{N}_{(1,1)}=\mathcal{N}_{(1,1)}^{\Psi},~0\leq\mathcal{N}_{(1,1)}^{\Psi}\leq 2~,\n  \\
\mathcal{N}_{(0,0)}=\mathcal{N}_{(0,0)}^{\Psi}+\mathcal{N}_{V},~0\leq
\mathcal{N}_{(0,0)}^{\Psi}\leq 2~,\n\label{(0,0)}
\Ieea
where $\mathcal{N}_{(2,0)}$ is the number of $(2,0)$-tensors, $\mathcal{N}_{(1,1)}$ 
is the number of real $(1,1)$-tensors, $\mathcal{N}_{(0,0)}$ is the number of real 
$(0,0)$-tensors and the above expressions give their decomposition to contributions coming from the 
various superfields.
For type (1), we have fields with two different dimensions so we get:
\Ibea{l}
\mathcal{K}_{[2]}={\cal K}^{\Psi}_{(2,0)}+{\cal K}^{\Psi}_{(1,1)}~,\n   \\
\mathcal{K}_{[1]}=k^{\Psi}_{(2,0)}+k^{\Psi}_{(1,1)}+k^{V}_{(1,1)}~,\n
\Ieea
where ${\cal K}  _{[2]}$ is the contribution of superfields to type (1) fields with 
engineering dimension two and similarly ${\cal K}  _{[1]}$ for dimension one. 
Because they come in pairs it must be true that:
\Ibea{l}
{\cal K}^{\Psi}_{(2,0)}=k^{\Psi}_{(1,1)}+k^{V}_{(1,1)}~,\n   \\
{\cal K}^{\Psi}_{(1,1)}=k^{\Psi}_{(2,0)}\n
\Ieea
and therefore 
\bea
{\cal K}  \equiv{\cal K}  _{[2]}={\cal K}  _{[1]}~.
\eea
Similarly with the type (2) contributions, all of the type (1) contributions have 
appropriate upper bounds which state that we can not have more than what it is provided 
by the superfields. Among these constraints, two are special. As we can see from the tables above
there can be only one $(2,0)$-tensor with dimensions $2$, therefore we must have
\Ibea{l}
{\cal N}^{\Psi}_{(2,0)}+{\cal K}^{\Psi}_{(2,0)}\leq 1\n\label{mix1}
\Ieea  
A similar argument applies to $(1,1)$-tensors with dimension 2, thus we get
\Ibea{l}
{\cal N}^{\Psi}_{(1,1)}+{\cal K}^{\Psi}_{(1,1)}\leq 2\n\label{mix2}
\Ieea
However, there is an important observation we should do. 
Out of all type (1), $(1,1)$-tensors with dimensions 1 we can generate, one of them
will play the role of the propagating spin 1. Because the spin 1 contribution to the total bosonic d.o.f has been taken into account by
(\ref{spin1}) we should not count it again.
This is a very special situation that arises only when we describe spin 1 propagating d.o.f in the presence of type (1) auxiliary fields.
That is because the type (1) auxiliary components have the same structure and gauge transformation law with the spin 1 component and for that reason we have to be cautious in order to guarantee the presence of an appropriate spin one components. Obviously, this is not going to be an issue for higher spins.
Putting everything together we get that
\bea
\mathcal{A}_{B}=6\mathcal{N}_{(2,0)}+4\mathcal{N}_{(1,1)}+\mathcal{N}_{(0,0)}+6{\cal K}
\eea
and therefore the condition of matching the bosonic d.o.f with the fermionic ones takes the form
\bea
3+\mathcal{A}_{B}=12+\mathcal{A}_{F}~.
\eea
This condition can be written in the following way
\Ibea{l}
6\left[{\cal N}^{\Psi}_{(2,0)}+{\cal K}^{\Psi}_{(2,0)}+{\cal K}^{\Psi}_{(1,1)}\right]+4{\cal N}^{\Psi}_{(1,1)}+{\cal N}^{\Psi}_{(0,0)}+{\cal N}_{V}=17~.\n\label{D1}
\Ieea
This is the Diophantine equation we have to solve together with all the various inequalities that constraint the value of the various coefficients.\\
\underline{Solutions}:\\
(\emph{i}) - First of all, we have to check whether there is a solution without the need for extra superfields, meaning ${\cal N}_{V}=0$.
In this case we have $k^{V}_{(1,1)}=0$ which also means that the physical spin 1 d.o.f will come out of the real or imaginary part of $\Psi^{(0,1)(S)}_{\a\ad}$. According to Table 3, this can happen if $\Psi_{\a}$ has a transformation that includes the term $\delta\Psi_{\a}=\D_{\a}K$, where $K$ is either real or imaginary. The consequence is that the other part of $\Psi^{(0,1)(S)}_{\a\ad}$ can be eliminated and therefore we get the conditions
\Ibea{l}
{\cal K}^{\Psi}_{(2,0)}=k^{\Psi}_{(1,1)}=0~.\n
\Ieea
Wit all the above in mind and due to the inequalities (\ref{mix1}, \ref{mix2}, \ref{(0,0)}) we can prove that there exist a unique solution
\Ibea{l}
{\cal N}^{\Psi}_{(2,0)}=1,~{\cal N}^{\Psi}_{(1,1)}=1,~{\cal N}^{\Psi}_{(0,0)}=1,~{\cal K}^{\Psi}_{(1,1)}=1,~k^{\Psi}_{(2,0)}=1~.\n
\Ieea
This solution correspondes to the Ogievetsky - Sokatchev description of the ($\tfrac{3}{2}, 1$) multiplet \cite{OgSok}
and the off-shell spectrum of the theory is:
\begin{enumerate}
\item[1.] one $(2,0)$-tensor of type (2): $t_{\a\b},~\delta_{g}t_{\a\b}=0,~[t_{\a\b}]=2$
\item[2.] one real $(1,1)$-tensor of type (2): $A_{\a\ad},~A_{\a\ad}=\bar{A}_{\a\ad},~\delta_{g}A_{\a\ad}=0,~[A_{\a\ad}]=2$
\item[3.] one real $(0,0)$-tensor of type (2): $P,~P=\bar{P},~\delta_{g}P=0,~[P]=2$
\item[4.] one real $(1,1)$-tensor of type (1): $V_{\a\ad},~V_{\a\ad}=\bar{V}_{\a\ad},~\delta_{g}V_{\a\ad}=\pa_{\a\ad}\lambda,~\lambda=\bar{\lambda},~[V_{\a\ad}]=2$
\item[5.] one $(2,0)$-tensor of type (1): $\omega_{\a\b},~\delta_{g}\omega_{\a\b}=i\pa_{(\b}{}^{\ad}\ell_{\a)\ad},~\ell_{\a\ad}=\bar{\ell}_{\a\ad}~,\ell_{\a\ad}\sim\ell_{\a\ad}+\pa_{\a\ad}\ell,~\ell=\bar{\ell},~[\omega_{\a\b}]=1$
\end{enumerate}
(\emph{ii}) - Let us investigate the existence of solutions with ${\cal N}_{V}\neq 0$. Following the requirement of economy, we will assume that ${\cal N}_V$ takes the least possible value and if this is not enough to solve (\ref{D1}) then we will increase its value by one and check again. So for ${\cal N}_V=1$ we get:
\Ibea{l}
6\left[{\cal N}^{\Psi}_{(2,0)}+{\cal K}^{\Psi}_{(2,0)}+{\cal K}^{\Psi}_{(1,1)}\right]+4{\cal N}^{\Psi}_{(1,1)}+{\cal N}^{\Psi}_{(0,0)}=16~,\n\\
{\cal N}^{\Psi}_{(2,0)}+{\cal K}^{\Psi}_{(2,0)}\leq 1~,\n\\
{\cal N}^{\Psi}_{(1,1)}+{\cal K}^{\Psi}_{(1,1)}\leq 2~,\n\\
{\cal N}^{\Psi}_{(0,0)}\leq 2~.\n
\Ieea
This system has two solutions:
\Ibea{l}
(\a)~~~~{\cal N}^{\Psi}_{(2,0)}+{\cal K}^{\Psi}_{(2,0)}=1,~{\cal K}^{\Psi}_{(1,1)}=1,~{\cal N}^{\Psi}_{(1,1)}=1,~{\cal N}^{\Psi}_{(0,0)}=0~,\n\label{sol1}\\
(\b)~~~~{\cal N}^{\Psi}_{(2,0)}+{\cal K}^{\Psi}_{(2,0)}=1,~{\cal K}^{\Psi}_{(1,1)}=0,~{\cal N}^{\Psi}_{(1,1)}=2,~{\cal N}^{\Psi}_{(0,0)}=2~.\n\label{sol2}
\Ieea
However solution $(\a)$, due to ${\cal N}^{\Psi}_{(0,0)}=0$, demands part of the transformation of $\Psi_{\a}$ to include a term $\D_{\a}L$ for an unconstrained $L$. The result is that component $\Psi^{(0,1)(S)}_{\a\ad}$ can be completely removed ($k^{\Psi}_{(1,1)}=0$) and therefore the spin 1 description must come from $V^{(1,1)}_{\a\ad}$ ($k^{V}_{(1,1)}=0$). Hence we must have ${\cal K}^{\Psi}_{(2,0)}=0$. The result is that the off-shell spectrum for solution ($\a$) is identical to that of case ($\emph{i}$) with ${\cal N}_V=0$. So we get that same theory using one extra superfield and for that reason this solution is rejected. The lesson here is that it is important to develop a method that will help us filter out equivalent theories and give only the different descriptions. More on that in the following section.

Solution $(\b)$ is a genuine different off-shell description of the same multiplet with a different spectrum. Using similar argument as before, we can show that ${\cal K}^{\Psi}_{(2,0)}=0$ and therefore the list of off-shell auxiliary components required are:
\begin{enumerate}
\item[1.] one $(2,0)$-tensor of type (2): $t_{\a\b},~\delta_{g}t_{\a\b}=0,~[t_{\a\b}]=2$
\item[2.] two real $(1,1)$-tensors of type (2): $A_{\a\ad},~A_{\a\ad}=\bar{A}_{\a\ad},~\delta_{g}A_{\a\ad}=0,~[A_{\a\ad}]=2$\\
\vphantom{1}\hspace{34.5ex} $U_{\a\ad},~U_{\a\ad}=\bar{U}_{\a\ad},~\delta_{g}U_{\a\ad}=0,~[U_{\a\ad}]=2$
\item[3.] three real $(0,0)$-tensor of type (2): $P,~P=\bar{P},~\delta_{g}P=0,~[P]=2$\\
\vphantom{1}\hspace{34.5ex} $S,~S=\bar{S},~\delta_{g}S=0,~[S]=2$\\
\vphantom{1}\hspace{34.5ex} $L,~L=\bar{L},~\delta_{g}L=0,~[L]=2$
\end{enumerate}
This solution corresponds to the de Wit-van Holten description of the ($\tfrac{3}{2},1$) multiplet \cite{MGM1,MGM2} and in contrast to the ${\cal N}_{V}=0$ case, there are no type (1) bosonic components.

4 - The gauge transformation of the superfields for the two descriptions $(\emph{i})$ and $(\emph{ii})(\b)$ can be found in straight forward manner by looking at tables 2 and 3 and demanding the non participating components to be gauge removed and the components that describe the spin d.o.f to have the proper transformation laws. For case $(\emph{i})$ we showed that we need to have a term $\D_{\a}K$, where $K$ is either real or imaginary in order to make sure of the presence of spin one. Also, because we must have ${\cal N}^{\Psi}_{(0,0)}=1$, either the real or the imaginary part of component $\Psi^{(1,0)(A)}$ must be gauged away without removing the other one. It is obvious from table 3, that this is possible if part of the transformation law of $\Psi_{\a}$ is $\delta\Psi_{\a}=\Dd^2\D_{\a}\Lambda$ with $\Lambda$ either real or purely imaginary. Hence putting these two together we get
\bea
\delta\Psi_{\a}=\D_{\a}K_1+i\Dd^2\D_{\a}K_2~,~~K_{i}=\bar{K}_{i}~,
\eea
For solution $(\emph{ii})(\b)$ removing all the uneccessary components give
\bea
\delta\Psi_{\a}=\D^2K_{\a}+\Dd^2K_{\a}~,\\
\delta V=\D^{\a}K_{\a}+\Dd^{\ad}\bar{K}_{\ad}~.
\eea
%%%%%%%%%%%%%%%%%%%%%%%%%%%%%%%%%%%%%%
%%%%%%%%%%%%%%%%%%%%%%%%%%%%%%%%%%%%%%
%%%%%%%% - Supergravity multiplet - %%%%%%%%
\section{$Y=3/2$ - Supergravity supermultiplets}
We repeat the steps for the supergravity supermultiplet:\vspace{1ex}\\
1 - The dynamics are that of spin $2$ and $3/2$, therefore the corresponding off-shell d.o.f. are
\Ibea{l}
\text{spin}~2 : (s^2+2)|_{s=2} = 6~,\\
\text{spin}~3/2 : (4s^2+4s+4)|_{s=1} = 12~.
\Ieea
\vspace{-3ex}\\
2 - The highest spin is $2$, therefore the main superfield must be a real $(1,1)$-tensor $H_{
\a\ad}$ of zero dimensions which provides a real vector auxiliary boson which could be of type 
(1) or of type (2). We also know that $H_{\a\ad}$ must have a gauge transformation of type 
$\mathcal{I}$. Furthermore, all of its components with engineering dimensions less or equal 
than $\tfrac{1}{2}$ must be able to be gauged away. That means the components $H^{
(1,0)(S)}$, $H^{(1,0)(A)}$, $H^{(0,1)(S)}$, $H^{(0,1)(A)}$ must have algebraic terms in their 
transformations, hence  according to table 2 we note the gauge transformation of $H_{
\a\ad}$ must be
\bea
\delta_{g}H_{\a\ad}={\rm D}{}_{\a}\bar{L}_{\ad}-{\Bar {\rm D}}_{\ad}L_{\a}  \label{dH}
\eea
with $L_{\a}$ unconstrained.\vspace{1ex}\\
3 - Checking the matching of bosonic and fermionic d.o.f.\\
If the only auxiliary field is $H^{(2,2)}_{\a\ad}$, then it must be of type (2), hence $\mathcal{
A}_{B}=4,~\mathcal{A}_{F}=0$. However, because
\bea
6+4\neq 12+0
\eea
this can not be the case and we have to introduce extra fields. This means that in the 
superspace description there must be more superfields besides the main one. Respecting 
the uniqueness of the highest spin, these extra superfields must be lower rank tensors, 
thus the only options we have are $\mathcal{N}_{\chi}$ copies of a fermionic superfield 
$\chi_{\a}$ with dimension $\tfrac{1}{2}$, $\mathcal{N}_{\psi}$ copies of a fermionic 
superfield $\psi_{\a}$ with dimension $-\tfrac{1}{2}$ and $\mathcal{N}_{V}$ copies of 
a real bosonic superfield $V$ with dimension $0$. Let us analyze the various auxiliary 
fields they can contribute:
\vspace{-2ex}
\begin{center}
\begin{tabular}{l r l c c c}
$1~\times~H_{\a\ad}~,~[H_{\a\ad}]=0$:~& & Tensor & Dimensions & off-shell d.o.f. & multiplicity \\
& Bosons: & $(1,1)$ & [$2$] & $4$ [type (2)] & $\le1$\\
& & & [$2$] & $3$ [type (1)] & $\le1$\\ 
{}\\
\hline\hline
{}\\
$\mathcal{N}_{\chi}~\times~\chi_{\a}~,~[\chi_{\a}]=\tfrac{1}{2}$:~&
Fermions: & $(1,0)$ & [$\tfrac{5}{2}$] & $4$ & ${\cal N}_{\chi}$\\
{}\\
 & Bosons: & $(2,0)$ & [$2$] & $6$ [type (2)] & $\le{\cal N}_{\chi}$ \\
 &  &  & [$2$] & $3$ [type (1)] & $\le{\cal N}_{\chi}$\\
 &  &  & [$1$] & $3$ [type (1)] & $\le{\cal N}_{\chi}$\\
 &  & $(0,0)$ & [$2$] & $1$ [type (2)] & $\le 2{\cal N}_{\chi}$\\
 &  & $(1,1)$ & [$2$] & $4$ [type (2)] & $\le 2{\cal N}_{\chi}$ \\
 &  &  & [$2$] & $3$ [type (1)] & $\le 2{\cal N}_{\chi}$ \\
 &  &  & [$1$] & $3$ [type (1)] & $\le 2{\cal N}_{\chi}$ \\ 
{}\\
\hline\hline
{}\\
$\mathcal{N}_{\psi}~\times~\psi_{\a}~,~[\psi_{\a}]=-\tfrac{1}{2}$:~& Bosons: & $(2,0)$ 
& [$1$] & $3$ [type (1)] & $\le{\cal N}_{\psi}$ \\
& & $(1,1)$ & [$1$] & $3$ [type (1)] & $\le{\cal N}_{\psi}$ \\ 
{}\\
\hline\hline
{}\\
$\mathcal{N}_{V}~\times~V~,~[V]=0$:~& Bosons: & $(0,0)$ & [$2$] & $1$ [type (2)] & ${\cal N}_{V}$\\
& & $(1,1)$ & [$1$] & $3$ [type (1)] & $\le{\cal N}_{V}$
\end{tabular}
\end{center}
For fermions, there is a non-removable $(1,0)$ component coming from every copy 
of $\chi_{\a}$ with the correct dimensions and because fermionic auxiliary fields 
come in pairs, we can immediately write that
\bea
\mathcal{A}_{F}=8\mathcal{N}_{\chi}~.
\eea
For bosons, we have to consider both types (1) and (2). Potential type (2) contributions 
are
\Ibea{l} \n \label{N}
\mathcal{N}_{(2,0)}={\cal N}_{(2,0)}^{\chi}~,~0\leq{\cal N}_{(2,0)}^{\chi}\leq\mathcal{
N}_{\chi}~, \sn \\
\mathcal{N}_{(1,1)}=\mathcal{N}_{(1,1)}^{\chi}+\mathcal{N}_{(1,1)}^{H}~,~0\leq
\mathcal{N}_{(1,1)}^{\chi}\leq 2\mathcal{N}_{\chi}~,~0\leq\mathcal{N}_{(1,1)}^{H}\leq 
1~, \sn \\
\mathcal{N}_{(0,0)}=\mathcal{N}_{(0,0)}^{\chi}+\mathcal{N}_{(0,0)}^{V}~,~0\leq
\mathcal{N}_{(0,0)}^{\chi}\leq 2\mathcal{N}_{\chi}~,~\mathcal{N}_{(0,0)}^{V}=
\mathcal{N}_{V}~.  \sn
\Ieea
For the type (1) auxiliary bosons we get: 
\Ibea{l} 
{\cal K}  _{[2]}={\cal K}  ^{H}_{(1,1)}+{\cal K}  ^{\chi}_{(1,1)}+{\cal K}  ^{\chi}_{(2,0)}~,  \n\\
{\cal K}  _{[1]}=k^{\chi}_{(1,1)}+k^{\chi}_{(2,0)}+k^{\psi}_{(1,1)}+k^{\psi}_{(2,0)}+k^{V}_{(1,1)}~.  \n
\Ieea
Because they come in pairs it must be true that
\Ibea{l}\n\label{K1}
{\cal K}  ^{H}_{(1,1)}+{\cal K}  ^{\chi}_{(1,1)}=k^{\chi}_{(2,0)}+k^{\psi}_{(2,0)}~, \sn \\
{\cal K}  ^{\chi}_{(2,0)}=k^{\chi}_{(1,1)}+k^{\psi}_{(1,1)}+k^{V}_{(1,1)}~, \sn \\
{\cal K}  \equiv{\cal K}  _{[2]}={\cal K}  _{[1]}~.  \sn
\Ieea
Notice that $H_{\a\ad}$ contributes only in ${\cal K}  _{[2]}$. This is a consequence 
of its gauge transformation (\ref{dH}) which will remove any appropriate component
\footnote{Except the component $H^{(1,1)(A,A)}$ which is reserved to play the role of spin auxiliary component
and \\participate in the spin dynamics.}. Also 
$\psi_{\a}$ contributes only in ${\cal K}  _{[1]}$ because of its dimension. 
Similarly with type (2), all of these contributions have 
appropriate upper bounds which state that we can not have more than what it is provided 
by the superfields. Nevertheless component $H^{(2,2)}$ has a special status because 
no matter what it has to be present and that leads to the following condition
\bea
{\cal K}  ^{H}_{(1,1)}+\mathcal{N}_{(1,1)}^{H}=1~. \label{K2}
\eea
Putting everything together we are led to
\bea
\mathcal{A}_{B}=6\mathcal{N}_{(2,0)}+4\mathcal{N}_{(1,1)}+\mathcal{N}_{(0,0)}+6{\cal K}~.  
\eea
Now, we can state the demand for matching of bosons and fermions
\bea
6\mathcal{N}_{(2,0)}+4\mathcal{N}_{(1,1)}+\mathcal{N}_{(0,0)}+6{\cal K}=6+8\mathcal{N}_{\chi}~. \label{msg}
\eea
The goal is to solve this Diophantine equation together with (\ref{N}), (\ref{K1}) and 
(\ref{K2}). A solution will provide the answer to how many superfields we need to 
have $({\cal N}^{*}_{\chi},~{\cal N}^{*}_{\psi},~{\cal N}^{*}_{V})$ and the specific 
structure of the components. However, as we learned in the previous section, not all 
solutions are blindly accepted and we are interested in the most economical ones. 
If a solution results to the triplet $({\cal N}^{*}_{\chi},~{\cal N}^{*}_{\psi},~{\cal N}^{*}_{V})$
then we do not accept any other solution with 
$({\cal N}^{*}_{\chi}+\delta_{\chi},~{\cal N}^{*}_{\psi}+\delta_{\psi},~{\cal N}^{*}_{V}+\delta_{V})$
for some positive, integers $\delta_{\chi},~\delta_{\psi},~\delta_{V}$, because such a 
solution can be reduced to the original solution plus extra stuff. In this sense, we can 
distinguish among the \emph{irreducible} descriptions and the \emph{non-irreducible} 
ones. Our aim is to find all, different, irreducible, solutions.\\
\underline{Solutions}:\\
(\emph{i}) - A good starting point for economical solutions is minimizing the degrees 
of freedom ($12+8{\cal N}_{\chi}$). So let us start with ${\cal N}_{\chi}=0$. For that case, 
we must have ${\cal N}_{(2,0)}^{\chi}={\cal N}_{(1,1)}^{\chi}={\cal N}_{(0,0)}^{\chi}={\cal 
K}  ^{\chi}_{(1,1)}={\cal K}  ^{\chi}_{(2,0)}=k^{\chi}_{(1,1)}=k^{\chi}_{(2,0)}=k^{\psi}_{(1,1)
}=k^{V}_{(1,1)
}=0$ and we are led to the system of equations:
\Ibea{l} 
4{\cal N}^{H}_{(1,1)}+{\cal N}_{V}+6{\cal K}  ^{H}_{(1,1)}=6~, \n \\
{\cal K}  ^{H}_{(1,1)}+\mathcal{N}_{(1,1)}^{H}=1~,\n\\
{\cal K}  ^{H}_{(1,1)}=k^{\psi}_{(2,0)}\leq\mathcal{N}_{\psi}~,  \n\\
0\leq\mathcal{N}_{(1,1)}^{H}\leq 1~. \n
\Ieea
This system has exactly two solutions
\begin{enumerate}
\item[$\alpha )$]~${\cal N}^{H}_{(1,1)}=0~,~{\cal N}_{V}=0~,~{\cal K}  ^{H}_{(1,1)}=k^{
\psi}_{(2,0)}=1\leq{\cal N}_{\psi}$~,
\item[$\beta )$]~${\cal N}^{H}_{(1,1)}=1~,~{\cal N}_{V}=2~,~{\cal K}  ^{H}_{(1,1)}=k^{
\psi}_{(2,0)}=0\leq{\cal N}_{\psi}$~.
\end{enumerate}
The first one corresponds to the triplet $({\cal N}^{*}_{\chi}=0,~{\cal N}^{*}_{\psi}\geq 
1,~{\cal N}^{*}_{V}=0)$ and as explained previously, it gives the ``\emph{irreducible}'' 
configuration of $(0,1,0)$. This will turn out to be the $12\times 12$, new-minimal 
formulation of supergravity supermultiplet with a superspace description based on 
$H_{\a\ad}$ and $\psi_{\a}$. The superfield $\psi_{\a},~[\psi_{\a}]=-\tfrac{1}{2}$ that 
appears here is the prepotential of the real linear superfield $[~U={\rm D}{}^{\a}{\Bar 
{\rm D}}^2\psi_{\a}+{\Bar {\rm D}}^{\ad}{\rm D}{}^2\bar{\psi}_{\ad}~]$ that is usually 
used to describe new-minimal supergravity. The off-shell auxiliary component 
spectrum has:
\begin{enumerate}
\item[1.] one real $(1,1)$-tensor of type (1): $A_{\a\ad},~A_{\a\ad}=\bar{A}_{\a\ad},~\delta_g A_{\a\ad}=\pa_{\a\ad}\lambda,~\lambda=\bar{\lambda},~[A_{\a\ad}]=2$
\item[2.] one $(2,0)$-tensor of type (1): $\omega_{\a\b},~\delta_g \omega_{\a\b}=i\pa_{(\b}{}^{\ad}\ell_{\a)\ad},~\ell_{\a\ad}=\bar{\ell}_{\a\ad},~\ell_{\a\ad}\sim\ell_{\a\ad}+\pa_{\a\ad}\ell,~\ell=\bar{\ell},~[\omega_{\a\b}]=1$
\end{enumerate}
We can also see that by changing the auxiliary vector from real to imaginary, one can get a slightly
different formulation of the supergravity supermultiplet which corresponds to new-new-minimal supergravity
which is using a different real linear superfield $U=i(\D^{\a}\Dd^2\psi_{\a}-\Dd^{\ad}\D^2\bar{\psi}_{\ad})$.\\
{}\\
The second solution gives the irreducible triplet $(0,0,
2)$. This corresponds to the $12\times 12$, old minimal description of the supergravity 
supermultiplet based on $H_{\a\ad}$ and two real superfields $V_1,~V_2$, which can be 
combined into a complex scalar superfield  which is the prepotential of the chiral 
superfield $[~\Phi={\Bar {\rm D}}^2(V_1+iV_2)~]$ that is usually used in the old-minimal 
formulation. The off-shell auxiliary spectrum for this case is, as expected,:
\begin{enumerate}
\item[1.] one real $(1,1)$-tensor of type (2): $A_{\a\ad},~A_{\a\ad}=\bar{A}_{\a\ad},~\delta_{g}A_{\a\ad}=0,~[A_{\a\ad}]=2$
\item[2.] two real $(0,0)$-tensor of type (2): $S,~S=\bar{S},~\delta_{g}S=0,~[S]=2$\\
{\vphantom{1}}\hspace{33.2ex} $P,~P=\bar{P},~\delta_{g}P=0,~[P]=2$
\end{enumerate}
\vspace{2ex}
(\emph{ii}) - Now, we consider solutions with ${\cal N}_{\chi}\neq0$. 
Because we have already found solutions that correspond to the triplets $(0,1,0)$ and $(0,0,2)$, 
the solutions with ${\cal N}_{\chi}\neq0$ in order to be acceptable as \emph{irreducible} 
must have ${\cal N}_{\psi}< 1 \Rightarrow {\cal N}_{\psi}=0$ and ${\cal N}_{V}< 2$.
For reasons of economy we do incremental steps so let's consider first the case of ${\cal N}_{\chi}=1$.
\bea
6\left[{\cal N}_{(2,0)}^{\chi}+{\cal K}  ^{\chi}_{(2,0)}+{\cal K}  ^{\chi}_{(1,1)}\right]+4{\cal 
N}^{\chi}_{(1,1)}+2{\cal K}  ^{H}_{(1,1)}+{\cal N}^{\chi}_{(0,0)}+{\cal N}_V=10\label{sg2}
\eea
with
\Ibea{l}\n 
0\leq\mathcal{N}_{(2,0)}^{\chi}+{\cal K}  ^{\chi}_{(2,0)}\leq1~,\sn\\
0\leq\mathcal{N}_{(1,1)}^{\chi}+{\cal K}  ^{\chi}_{(1,1)}\leq2~,\sn\\
0\leq\mathcal{N}_{(0,0)}^{\chi}\leq2~,~0\leq\mathcal{N}_V\leq1~,\sn\\ 
0\leq k^{\chi}_{(2,0)}={\cal K}  ^{H}_{(1,1)}+{\cal K}  ^{\chi}_{(1,1)}\leq1~,\sn\\
0\leq {\cal K}  ^{\chi}_{(2,0)}=k^{\chi}_{(1,1)}+k^{V}_{(1,1)}\leq1 ~.\sn
\Ieea
On top of the above, we have a set of conditional constraints coming from table 4. In this case these constraints take the form:
\begin{enumerate}
\item If ${\cal N}^{\chi}_{(2,0)}+{\cal K}^{\chi}_{(2,0)}=0$, then $k^{\chi}_{(2,0)}=k^{\chi}_{(1,1)}=0$.
\item If ${\cal N}^{\chi}_{(1,1)}+{\cal K}^{\chi}_{(1,1)}\leq 1$, then $k^{\chi}_{(2,0)}=k^{\chi}_{(1,1)}=0$.
\item We can not have ${\cal N}^{\chi}_{(2,0)}\leq 1$ and ${\cal N}^{\chi}_{(1,1)}+{\cal K}^{\chi}_{(1,1)}\leq 1$ because there will be no candidate for the
second fermionic auxiliary component needed to complete the pair.
\end{enumerate}
Due to all the above constraints there is a unique solution of (\ref{sg2})
\Ibea{l}\n
{\cal N}^{\chi}_{(0,0)}+{\cal N}_V=2~,\\
{\cal N}_{(2,0)}^{\chi}+{\cal K}  ^{\chi}_{(2,0)}+{\cal K}  ^{\chi}_{(1,1)}=0~ \to~{\cal N}_{(2,0)}^{\chi}={\cal K}  ^{\chi}_{(2,0)}={\cal K}  ^{\chi}_{(1,1)}=k^{\chi}_{(2,0)}=k^{V}_{(1,1)}=0~,\\
{\cal N}^{\chi}_{(1,1)}=2~,\\
{\cal K}  ^{H}_{(1,1)}=0~.
\Ieea
This solution stands for any value of ${\cal N}_{V}<2$, hence the irreducible piece will correspond to ${\cal N}_{V}=0$ giving the triplet $(1,0,0)$.
This 
is the $20\times 20$, non-minimal formulation of supergravity with a superspace description 
based on $H_{\a\ad}$ and $\chi_{\a}$ which is the prepotential of the complex linear 
superfield $[~\Gamma={\Bar {\rm D}}^{\ad}\bar{\chi}_{\ad}~]$ which traditionally used.
The off-shell auxiliary spectrum of the theory has:
\begin{enumerate}
\item[1.] three real $(1,1)$-tensors of type (2): $A_{\a\ad},~A_{\a\ad}=\bar{A}_{\a\ad},~\delta_{g} A_{\a\ad}=0,~[A_{\a\ad}]=2$\\
{\vphantom{1}}\hspace{36ex} $u_{\a\ad},~u_{\a\ad}=\bar{u}_{\a\ad},~\delta_{g} u_{\a\ad}=0,~[u_{\a\ad}]=2$\\
{\vphantom{1}}\hspace{36ex} $v_{\a\ad},~v_{\a\ad}=\bar{v}_{\a\ad},~\delta_{g} v_{\a\ad}=0,~[v_{\a\ad}]=2$
\item[2.] two real $(0,0)$-tensors of type (2): $S,~S=\bar{S},~\delta_{g} S=0,~[S]=2$\\
{\vphantom{1}}\hspace{34ex} $P,~P=\bar{P},~\delta_{g} P=0,~[P]=2$
\item[3.] two $(1,0)$-tensors: $\b_{\a},~\delta_{g} \b_{\a}=0,~[\b_{\a}]=\tfrac{5}{2}$\\
{\vphantom{1}}\hspace{18ex} $\rho_{\a},~\delta_{g} \rho_{\a}=0,~[\rho_{\a}]=\tfrac{3}{2}$
\end{enumerate}
4 - Based on the Diophantine equation we managed to classify and generate all possible 
irreducible formulation of supergravity supermultiplet. The answer was the familiar minimal 
(old and new) and non-minimal formulations. To complete the discussion, we have to find 
the gauge transformations for the superfields. Of course we know the gauge transformation 
of $H_{\a\ad}$ but we have to find it for $\chi_{\a},~\psi_{\a}$ and $V$ in a way that is 
consistent with the component spectrum.

For the old-minimal solution $(0,0,2)$, we combine the two real scalars $V_1,~V_2$ into 
one complex superfield $W$ and we demand that its gauge transformation is such that it 
removes all its components (except of course $W^{(2,2)}$ which can not be removed). 
Looking through Table 3, we find that in order to remove $W^{(1,2)(S)}$ we must have a 
${\rm D}{}^{\a}J_{\a}$ term in the transformation with $J_{\a}$ unconstrained. Similarly in 
order to remove $W^{(2,1)(S)}$ we must have a ${\Bar {\rm D}}^{\ad}I_{\ad}$ term in the 
transformation with $I_{\ad}$ unconstrained. Also, we observe that these two terms are 
enough to gauge away all other components. Furthermore, in order to use $H_{\a\ad}$ 
and $W$ to construct a non-trivial gauge invariant superspace theory we must have 
the set of gauge parameters of $W$ and the set of 
gauge parameters of $H_{\a\ad}$ to have an non-empty overlap.
Hence $J_{\a}$ must be identified with $L_{\a}$. So the gauge transfornmation of $W$ is
\bea
\delta_g W= {\rm D}{}^{\a}L_{\a}+{\Bar {\rm D}}^{\ad}\Lambda_{\ad}    \label{dW}
\eea
for an arbitrary $\Lambda_{\ad}$. As we said previously, $W$ must be interpreted as the 
prepotential of the chiral superfield $\Phi$ $[\Phi={\Bar {\rm D}}^2 W]$ that is usually used. 
This is in agreement with (\ref{dW}) because it produces the correct transformation for 
$\Phi$ $[\delta\Phi={\Bar {\rm D}}^2 {\rm D}{}^{\a}L_{\a}]$.

For the new-minimal solution $(0,1,0)$, things become a little more interesting. The $\psi^{
(1,2)(S)}_{\a\b}$ component of superfield  $\psi_{\a}$ must survive the gauge transformation 
because it generates the required $(2,0)$-tensor of type (1). In section (\ref{type1}) we showed 
that these type of components have a very special transformation. So the transformation of 
$\psi_{\a}$ must be choosen in a way that it respects these properties and also removes all 
other components. Going through Table 3, we find that in order to remove components $
\psi^{(2,1)(S)}_{\a\ad}$ we must have a term ${\Bar {\rm D}}^{\ad}I_{\a\ad}$ in the transformation 
of $\psi_{\a}$, with $I_{\a\ad}$ unconstrained. Moreover, in order to remove component $\psi^{
(1,2)(A)}_{}$ we must have a term ${\rm D}{}_{\a}L$, with $L$ unconstrained. These two terms 
are enough to remove all other components of $\psi_{\a}$, but when we check the transformation 
of $\psi^{(1,2)(S)}_{\a\b}$ we find that it does not have the correct properties. Specifically the 
gauge parameter, which in this case is $i[{\rm D}{}_{\a},{\Bar {\rm D}}_{\ad}]L|$ is not real. In 
order to make it real we must choose $L$ to be imaginary, but in return this choice conflicts 
with the gauge removal of $\psi^{(1,2)(A)}$. So it seems there is no consistent choice for 
the transformation of $\psi_{\a}$. However, even if we ignore this issue, we are still left with the 
fact that neither gauge parameters $(L,~I_{\a\b})$ has the structure of the gauge parameter of 
$H_{\a\ad}$, so how can we use both these fields to construct an invariant superspace theory? 
Well, there is a way out that solves both issues at the same time and it is hidden in a subtle 
details.

In equation (\ref{K1}) we claimed that superfield $H_{\a\ad}$ does not contribute in the ${\cal 
K}  _{[1]}$ terms, because we have used its gauge transformation to remove such terms. That 
means, we can now use the gauge parameter $L_{\a}$ and add it algebraically to the 
transformation of $\psi_{\a}$ without risking that it will remove the entire superfield. This allow 
us to choose $L$ imaginary, remove $\psi^{(1,2)(A)}$ and not remove $\psi^{(1,2)(S)}_{\a\b}$ 
because the relevant component $L^{(1,2)(S)}$ has already been used to remove the component
$H^{(1,1)(S,A)}_{\a\b}$. So, the result is the gauge transformation of $\psi_{\a}$ must be
\bea
\delta_{g}\psi_{\a}=L_{\a}+{\rm D}{}_{\a}K+{\Bar {\rm D}}^{\ad}\Lambda_{\a\ad}~,~K=-\bar{K}~.
\eea
This also agrees with the interpretation of $\psi_{\a}$ as the prepotential of a real linear $U$, 
because it generates the known transformation for it $\delta U={\rm D}{}^{\a}{\Bar {\rm D}}^2L_{
\a}+c.c.$~\footnote{Very similar arguments will give rise to the gauge transformation law for the 
new-new-minimal description of supergravity supermultiplet, where $K$ is real}.

Finally, for the non-minimal solution $(1,0,0)$, we must have a tranformation that does not eliminate 
the components of $\chi^{(2,1)(S)}_{\a\ad}$, $\chi^{(1,2)(A)}$. Also we must have one $(1,0)$-tensor 
fermion with dimensions $\tfrac{3}{2}$ which must be there in order together with $\chi^{(2,2)}$ to 
complete the pair of auxiliary fermions. All other components must be removed by the gauge  
transformation. The removal of $\chi^{(1,2)(S)}_{\a\b}$ forces us to have a term like ${\rm D}{}^{
\b}J_{\a\b}$ in the transformation law of $\chi_{\a}$ with $J_{\a\b}$ unconstrained. On the other 
hand, the removal of $\chi^{(1,0)(A)}_{}$ forces us to introduce a term ${\Bar {\rm D}}^2\Lambda_{
\a}$ with arbitrary $\Lambda_{\a}$ or a term ${\rm D}{}_{\a}L$ with $L$ constrained such that it 
does not remove $\chi^{(1,2)(A)}$ or a term ${\Bar {\rm D}}^{\ad}I_{\a\ad}$ with $I_{\a\ad}$ 
constrained such that it does not remove $\chi^{(2,1)(S)}_{\a\ad}$. At this point, the most 
general structure allowed is
\bea
\delta_{g}\chi_{\a}\sim{\rm D}{}^{\b}\Lambda_{\a\b}+{\Bar {\rm D}}^2\Lambda_{\a}+{\rm D}{}_{\a}{
\Bar {\rm D}}^{\ad}\bar{K}_{\ad}+{\Bar {\rm D}}^{\ad}{\rm D}{}_{\a}\bar{I}_{\ad}
\eea
with unconstrained parameters. The parameters $\Lambda_{\a},~K_{\a},~I_{\a}$ eventually have 
to be identified with the $L_{\a}$, which means the last term can be removed by a redefinition 
of $\chi_{\a}$ $[\chi_{\a}\to\chi_{\a}+{\Bar {\rm D}}^{\ad}H_{\a\ad}]$. Also, there are two fermions 
left $\chi^{(2,0)}_{\a},~\chi^{(1,1)(A,S)}_{\ad}$ and we have to remove only one of them. We can 
either use the second term and remove the first fermion or use the third term and remove the first 
fermion. So the expression for the gauge transformation of $\chi_{\a}$ is:
\bea
\delta_{g}\chi_{\a}={\Bar {\rm D}}^2L_{\a}+\tfrac{f}{2}{\rm D}{}_{\a}{\Bar {\rm D}}^{\ad}\bar{L}_{
\ad}+{\rm D}{}^{\b}\Lambda_{\a\b} 
\eea
where the parameter $f$ controls the relative coefficient between the first two terms. The interpretation 
of $\chi_{\a}$ as the prepotential of the complex linear $\Gamma$ compensator that is usually used, 
is in agreement with this result because it generates the correct transformation for it $\delta_{g}\Gamma
={\Bar {\rm D}}^{\ad}{\rm D}{}^2\bar{L}_{\ad}+f{\Bar {\rm D}}^2{\rm D}{}^{\a}L_{\a}$. However, we 
observe that if $f$ becomes very large then the second term in the transformation of $\Gamma$ 
dominates and $\delta\Gamma$ reduces to the transformation of the chiral in the first solution. 
Therefore, we want to impose the constraint $f\neq\infty$. Also, we can redefine $\Gamma$ to $
\tilde{\Gamma}=\Gamma-f{\Bar {\rm D}}^{\ad}{\rm D}{}^{\a}H_{\a\ad}$ and show that
\bea
\delta_g\tilde{\Gamma}=(1-2f){\Bar {\rm D}}^{\ad}{\rm D}{}^2\bar{L}_{\ad}-f{\rm D}{}^{\a}{\Bar {\rm 
D}}^2L_{\a}~.
\eea

Hence, in order not to make contact with the second solution we must have $f\neq1$ and $f\neq
\tfrac{1}{3}$, thus giving as final result
\bea
\delta_{g}\chi_{\a}={\Bar {\rm D}}^2L_{\a}+\tfrac{f}{2}{\rm D}{}_{\a}{\Bar {\rm D}}^{\ad}\bar{L}_{\ad}
+{\rm D}{}^{\b}\Lambda_{\a\b}~,~f\neq 1,\tfrac{1}{3}, \infty~.
\eea

Having the set of superfields that we should use, their gauge transformation laws and knowing 
that the theory exist determines completely the superspace and component action for all 
formulations of a $Y=\tfrac{3}{2}$ theory.

This completes the application of this approach to the supergravity supermultiplet. We managed 
to show that it is possible to classify and derive the off-shell component spectrum from very basic 
requirements without constructing the action of the theory. This is the exact opposite of the 
mainstream approach where the spectrum of the theory is been derived from the details of 
an action.
%%%%%%%%%%%%%%%%%%%%%%%%%%%%%%%%%%%%%%
%%%%%%%%%%%%%%%%%%%%%%%%%%%%%%%%%%%%%%
%%%%% - Higher Half-integer spin - %%%%
\section{Arbitrary Half-integer superspin}
We can continue applying successfully this method to even higher 
superspin theories. In this section, we will attempt to do so for the cases of arbitrary half-integer 
superspin theories. We will find that going beyond the supergravity multiplet introduces an 
interesting twist that can be exploited in order to successfully solve the Diophantine equation and obtain the answer.\vspace{1.5ex}\\
1 - Consider the $Y=s+\tfrac{1}{2}$ multiplet. The dynamics of the theory will be described 
by one spin $s+1$ and one spin $s+\tfrac{1}{2}$. Therefore the off-shell degrees of freedom 
they provide are
\Ibea{l}
\text{spin}~s+1 : (j^2+2)|_{j=s+1} = s^2+2s+3~,\\
\text{spin}~s+\tfrac{1}{2} : (4j^2+4j+4)|_{j=s} = 4s^2+4s+4
\Ieea
and they will be described by the component fields $h_{\a(s+1)\ad(s+1)}$, $h_{\a(s-1)\ad(s-1)}$, 
$\psi_{\a(s+1)\ad(s)}$, $\psi_{\a(s)\ad(s-1)}$, and $\psi_{\a(s-1)\ad(s-2)}$ respectively together 
with appropriate gauge transformations.\\
{}\\
2 - The highest spin is $s+1$, therefore the main superfield in the superspace description 
must be a real scalar superfield $H_{\a(s)\ad(s)}$. Following the same argument as in the 
previous case we can uniquely determine its gauge transformation to be
\bea
\delta_{g}H_{\a(s)\ad(s)}=\tfrac{1}{s!}{\rm D}{}_{(\a_s}\bar{L}_{\a(s-1))\ad(s)}-\tfrac{1}{s!}{\Bar 
{\rm D}}{}  _{(\ad_s}L_{\a(s)\ad(s-1))}~.  \label{dHs}
\eea
{}\\
3 - Auxiliary superfields and matching bosons with fermions:\\
In order to find the list of appropriate auxiliary superfields, we can repeat the arguments we used in the previous sections,
and list all superfields with total number of indices less than $2s$. However for the general half-integer superspin supermultiplet
there is a quicker approach which is based on a 
qualitative difference with the lower spin supermultiplets such as supergravity multiplet. As we can see 
from Table 1, the lowest rank fermion that $H_{\a(s)\ad(s)}$ provides is a $(s,
s-1)$-tensor.  Therefore $H_{\a(s)\ad(s)}$ can not generate the $\psi_{\a(s-1)\ad(s-2)}$ component that is 
required for the off-shell description of spin $s+\tfrac{1}{2}$. For the case of supergravity 
multiplet, this issue was avoided because this component is not relevant for $s\leq1$.  So 
we know immediately, without checking the matching of bosons with fermions, that we 
need auxiliary superfields and unlike the previous cases, they have to provide not only 
auxiliary fields but also the missing dynamical component. The most economical way to generate
this missing component without introducing any other dynamical components with even lower rank is to consider
auxiliary superfields $A_{\a(k)\ad(l)}$ such that their lowest rank component $A^{(1,1)(A,A)}_{\a(k-1)\ad(l-1)}$
can be identified with $\psi_{\a(s-1)\ad(s-2)}$. This would suggest to consider the superfield
\Ibea{l}
k-1=s-1~,~ l-1=s-2~,~ d+1=\tfrac{3}{2} ~~~~\Rightarrow~~ \chi_{\a(s)\ad(s-1)}~,~d_{\chi}=\tfrac{1}{2}\n
\Ieea
However, we see in Table 2 that there is a possibility for 
component $A^{(1,1)(A,A)}_{\a(k-1)\ad(l-1)}$ to be gauged away by an appropriate gauge transformation. 
In that case, we go to the next lowest rank tensors within the same group. These are the components $A^{
(1,1)(S,A)}_{\a(k+1)\ad(l-1)}$,~ $A^{(1,1)(A,S)}_{\a(k-1)\ad(l+1)}$,~ $A^{(2,0)}_{\a(k)\ad(l)}$
,~ $A^{(0,2)}_{\a(k)\ad(l)}$\footnote{If $A^{(1,1)(A,A)}_{\a(k-1)\ad(l-1)}$ can be 
removed then so can $A^{(2,0)}_{\a(k)\ad(l)}$ or $A^{(0,2)}_{\a(k)\ad(l)}$.}. So, we must add to the list of potential
auxiliary superfields $\chi_{\a(s-1)\ad(s-2)}$, $\chi_{\a(s-2)\ad(s-3)}$.
This is a huge shortcut because instead of considering 
the entire list of potential superfields $\left\{A_{\a(s)\ad(s-1)}~,~\dots~,A\right\}$ we only have to consider
$\left\{\chi_{\a(s)\ad(s-1)}, \chi_{\a(s-1)\ad(s-2)}, \chi_{\a(s)\ad(s-3)}\right\}$.
Now, we analyze the set of potential auxiliary components these superfields introduce. 
First of all, due to the index structure all of the auxiliary bosons must be of type (2). 
Secondly, all of these superfields introduce auxiliary fermions. We obtain:
\begin{center}
\begin{tabular}{l r l c l c}
 & & Tensor & Dimensions & off-shell d.o.f. & multiplicity\\
$1~\times~H_{\a(s)\ad(s)}$:~& Bosons: & $(s, \, s)$ & [$2$] & $(s+1)^2$ & $1$ \\
\vspace{-2ex}\\
\hline\hline
\vspace{-2.0ex}\\
$\mathcal{N}_1~\times~\chi_{\a(s)\ad(s-1)}$:~&
Fermions: & $(s, \, s-1)$ & [$\tfrac{5}{2}$] & $2(s+1)s$ & ${\cal N}_{1}$ \\
\vspace{-1.5ex}\\
 & Bosons: & $(s+1,\, s-1)$ & [$2$] & $2(s+2)s$ & $\le{\cal N}_{1}$ \\
 &  & $(s-1,\, s-1)$ & [$2$] & $s^2$ & $\le2{\cal N}_{1}$\\
 &  & $(s, \, s)$ & [$2$] & $(s+1)^2$ & $\le2{\cal N}_{1}$ \\
 &  & $(s, \, s-2)$ & [$2$] & $2(s+1)(s-1)$ & $\le{\cal N}_{1}$\\
\vspace{-2ex}\\
\hline\hline
\vspace{-2.0ex}\\
$\mathcal{N}_2~\times~\chi_{\a(s-1)\ad(s-2)}$:~&
Fermions: & $(s-1,\, s-2)$ & [$\tfrac{5}{2}$] & $2s(s-1)$ & ${\cal N}_{2}$ \\
\vspace{-1.5ex}\\
 & Bosons: & $(s, \, s-2)$ & [$2$] & $2(s+1)(s-1)$ & $\le{\cal N}_{2}$ \\
 &  & $(s-2,\, s-2)$ & [$2$] & $(s-1)^2$ & $\le2{\cal N}_{2}$\\
 &  & $(s-1,\, s-1)$ & [$2$] & $s^2$ & $\le2{\cal N}_{2}$ \\
 &  & $(s-1,\, s-3)$ & [$2$] & $2s(s-2)$ & $\le{\cal N}_{2}$ \\
\vspace{-2ex}\\
\hline\hline
\vspace{-2.0ex}\\
$\mathcal{N}_3~\times~\chi_{\a(s)\ad(s-3)}$:~&
Fermions: & $(s,\, s-3)$ & [$\tfrac{5}{2}$] & $2(s+1)(s-2)$ & ${\cal N}_{3}$ \\
\vspace{-1.5ex}\\
 & Bosons: & $(s+1,\, s-3)$ & [$2$] & $2(s+2)(s-2)$ & $\le{\cal N}_{3}$ \\
 &  & $(s-1,\, s-3)$ & [$2$] & $2s(s-2)$ & $\le{\cal N}_{3}$\\
 &  & $(s,\, s-2)$ & [$2$] & $2(s+1)(s-1)$ & $\le{\cal N}_{3}$ \\
 &  & $(s,\, s-4)$ & [$2$] & $2(s+1)(s-3)$ & $\le{\cal N}_{3}$
\end{tabular}
\end{center} 
Therefore the auxiliary components coming from $H$ and the $\chi$s are
\Ibea{ll}\n
&\mathcal{A}_{B}=~(s+1)^2+2(s+2)s{\cal N}{}_{(s+1,\, s-1)}+2(s+1)(s-1){\cal N}_{(s, \, s-2)}+2s(s-2){\cal N}_{(s-1,\, s-3)} \sn\\
&~~~~~~~+2(s+2)(s-2){\cal N}_{(s+1,\, s-3)}+2(s+1)(s-3){\cal N}_{(s,\, s-4)}\\
&~~~~~~~+(s+1)^2{\cal N}_{(s, \, s)}+s^2{\cal N}_{(s-1,\, s-1)}+(s-1)^2{\cal N}_{(s-2,\, s-2
)}~,\\
{}\\
%%%
&\mathcal{A}_{F}=4(s+1)s{\cal N}_1+4s(s-1){\cal N}_2+4(s+1)(s-2){\cal N}_3\sn
\Ieea
where
\Ibea{l}\n
{\cal N}_{(s+1,\, s-1)}=N^{1}_{(s+1,\, s-1)}~~~~~~~~~~~~~~~~~~~~~~~~~~,~N^{1}_{(s+1,\, s-1)}\leq{\cal N}_1\\
%%%
{\cal N}_{(s, \, s-2)}=N^{1}_{(s, \, s-2)}+ N^{2}_{(s, \, s-2)}+N^{3}_{(s, \, s-2)}~~~,~N^{1}_{(s, \, s-2)}\leq{\cal 
N}_1~,~~N^{2}_{(s, \, s-2)}\leq{\cal N}_2~,~~N^{3}_{(s, \, s-2)}\leq{\cal N}_3\\
%%%
{\cal N}_{(s-1,\, s-3)}=N^{2}_{(s-1,\, s-3)}+N^{3}_{(s-1,\, s-3)}~~~~~~~~,~N^{2}_{(s-1,\, s-3)}
\leq{\cal N}_2~,~~N^{3}_{(s-1,\, s-3)}\leq{\cal N}_3~,\\
%%%
{\cal N}_{(s+1,\, s-3)}=N^{3}_{(s+1,\, s-3)}~~~~~~~~~~~~~~~~~~~~~~~~~,~N^{3}_{(s+1,\, s-3)}\leq{\cal N}_3~,\\
%%%
{\cal N}_{(s,\, s-4)}=N^{3}_{(s,\, s-4)}~~~~~~~~~~~~~~~~~~~~~~~~~~~~~~~,~N^{3}_{(s,\, s-4)}\leq{\cal N}_3~,\\
%%%
{\cal N}_{(s, \, s)}=N^{1}_{(s, \, s)}~~~~~~~~~~~~~~~~~~~~~~~~~~~~~~~~~~~~~,~N^{1}_{(s, \, s)}\leq 2{\cal N}_1~,\\
%%%
{\cal N}_{(s-1,\, s-1)}=N^{1}_{(s-1,\, s-1)}+N^{2}_{(s-1,\, s-1)}~~~~~~~~,~N^{1
}_{(s-1,\, s-1)}\leq 2{\cal N}_1~,~N^{2}_{(s-1,\, s-1)}\leq 2{\cal N}_2~,\\
%%%
{\cal N}_{(s-2,\, s-2)}=N^{2}_{(s-2,\, s-2)}~~~~~~~~~~~~~~~~~~~~~~~~~,~N^{2}_{(s-2,\, s-2)}\leq 2{\cal N}_2~.
\Ieea
The matching of the bosonic and fermionic degrees of freedom condition is
\bea
\underbrace{s^2+2s+3}_{\text{spin~}s+1}+\mathcal{A}_{B}=\underbrace{4s^2+4s+4}_{\text{spin~}s+\tfrac{1}{2}}+\mathcal{
A}_{F}~~
\Rightarrow~~~~~\mathcal{A}_{B}=2s^2+(s+1)^2+\mathcal{A}_{F}~.  \label{int-s-d}
\eea
This is the Diophantine equation we have to solve and the acceptable solutions 
will determine the spectrum of the higher spin supersymmetric theories. The brute 
force method of solving it is to interpret (\ref{int-s-d}) as a polynomial equation which 
must hold for every value of $s\geq2$ and therefore the coefficients for each power of $s$ 
in both sides must match. However, there is a more elegant method. 
First of all, notice that the $(s+1)^2$ term in the right hand side of the above equation will cancel exactly the $(s+1)^2$
term in $\mathcal{A}_{B}$, coming from the auxiliary d.o.f of superfield $H_{\a(s)\ad(s)}$.
The rest of the coefficients in $\mathcal{A}_{B}$ have the structure $(k)^2$ or $k(k-2)$ or $k(k-4)$ for some integers $k$.
Similarly, notice that the numerical coefficients in $\mathcal{A}_{F}$ are
of the form 
$(k+1)k$ or $(k+2)k$ for some integer $k$. So solving (\ref{int-s-d}) means that we must be able to 
convert from one structure to the other. This is indeed possible because of the following three identities
\Ibea{r,l}    \n\label{convert}
(I)~&4(k+m)k=2(k+m)^2+2(k+m)(k-m) ~,   \\
(II)~&4(k+m)k=2k^2+2(k+2m)k   ~,\\
(III)~&4(k+m)k=6k^2-2k(k-2m)~.
\Ieea
Using them, we convert all the terms that appear in $\mathcal{A}_{F}$:
\Ibea{l}\n\label{c1}
4(s+1)s{\cal N}_1=
\begin{cases}
2(s+1)^2{\cal N}_1+2(s+1)(s-1){\cal N}_1~,~(I)\\
2s^2{\cal N}_1+2(s+2)s{\cal N}_1~,~~~~~~~~~~~~~~~~(II)\\
6s^2{\cal N}_1-2s(s-2){\cal N}_1~,~~~~~~~~~~~~~~~~(III)
\end{cases}\\
\Ieea
\Ibea{l}\n\label{c2}
4s(s-1){\cal N}_2=
\begin{cases}
2s^2{\cal N}_2+2s(s-2){\cal N}_2~,~~~~~~~~~~~~~~~~(I)\\
2(s-1)^2{\cal N}_2+2(s+1)(s-1){\cal N}_2~,~(II)\\
6(s-1)^2{\cal N}_2-2(s-1)(s-3){\cal N}_2~,~(III)
\end{cases}
\Ieea
\Ibea{l}\n\label{c3}
4(s+1)(s-2){\cal N}_3=
\begin{cases}
2(s+1)^2{\cal N}_3+2(s+1)(s-5){\cal N}_3~,~~~(I)\\
2(s-2)^2{\cal N}_3+2(s+4)(s-2){\cal N}_3~,~~~(II)\\
6(s-2)^2{\cal N}_3-2(s-2)(s-8){\cal N}_3~,~~~(III)
\end{cases}
\Ieea
From the (\ref{c3}) we can easily see that ${\cal N}_3$ has to be zero, because non of the three identities can generate terms that appear in $\mathcal{A}_{B}$ and from (\ref{c1}, \ref{c2}) we deduce that only identities ($I$) and ($II$) are relevant. Applying ($I$) in (\ref{int-s-d})
we get a solution with ${\cal N}_{2}={\cal N}_{3}=0$,~
${\cal N}_{1}=1$:
\Ibea{ll}
N^{2}_{(\dots)}=0 ~,~&~ N^{1}_{(s+1,\, s-1)}=0~, \n\\
N^{3}_{(\dots)}=0 ~,~&~ N^{1}_{(s-1,\, s-1)}=2~, \n\\
 &~ N^{1}_{(s, \, s)}=2~, \n\\
 &~ N^{1}_{(s, \, s-2)}=1~. \n
\Ieea
Using ($II$) we get a solution with ${\cal N}_{1}={\cal N}_{3}=0$,
~${\cal N}_{2}=1$:
\Ibea{ll}
N^{1}_{(\dots)}=0 ~,~&~ N^{2}_{(s-1,\, s-1)}=2~, \n\\
N^{3}_{(\dots)}=0 ~,~&~ N^{2}_{(s, \, s-2)}=1~, \n\\
 &~ N^{2}_{(s-2,\, s-2)}=2~, \n\\
 &~ N^{2}_{(s-1,\, s-3)}=0~. \n
\Ieea
The result is that there are two different formulation of the theory because there are 
exactly two ways that we can convert the structure of the fermionic d.o.f. to bosonic d.o.f.
For the first solution, the theory 
has $8s^2+8s+4~\times~8s^2+8s+4$ d.o.f., its superspace description is based on 
the superfields $H_{\a(s)\ad(s)}$ and $\chi_{\a(s)\ad(s-1)}$ and the off-shell auxiliary 
component spectrum has:
\begin{enumerate}
\item[1.] three real $(s, \, s)$-tensors :\\
$A_{\a(s)\ad(s)},~A_{\a(s)\ad(s)}=\bar{A}_{\a(s)\ad(s)},~\delta_g A_{\a(s)\ad(s)}=0,~[A_{\a(s)\ad(s)}]=2$\\
$u_{\a(s)\ad(s)},~u_{\a(s)\ad(s)}=\bar{u}_{\a(s)\ad(s)},~\delta_g u_{\a(s)\ad(s)}=0,~[u_{\a(s)\ad(s)}]=2$\\
$v_{\a(s)\ad(s)},~v_{\a(s)\ad(s)}=\bar{v}_{\a(s)\ad(s)},~\delta_g v_{\a(s)\ad(s)}=0,~[v_{\a(s)\ad(s)}]=2$
\item[2.] two real $(s-1,s-1)$-tensors :\\
$S_{\a(s-1)\ad(s-1)},~S_{\a(s-1)\ad(s-1)}=\bar{S}_{\a(s-1)\ad(s-1)},~\delta_g S_{\a(s-1)\ad(s-1)}=0,~[S_{\a(s-1)\ad(s-1)}]=2$\\
$P_{\a(s-1)\ad(s-1)},~P_{\a(s-1)\ad(s-1)}=\bar{P}_{\a(s-1)\ad(s-1)},~\delta_g P_{\a(s-1)\ad(s-1)}=0,~[P_{\a(s-1)\ad(s-1)}]=2$
\item[3.] one $(s, \, s-2)$-tensor : $U_{\a(s)\ad(s-2)},~\delta_g U_{\a(s)\ad(s-2)}=0,~[U_{\a(s)\ad(s-2)}]=2$
\item[4.] two $(s, \, s-1)$-tensors : $\b_{\a(s)\ad(s-1)},~\delta_g \b_{\a(s)\ad(s-1)}=0,~[\b_{\a(s)\ad(s-1)}]=\tfrac{5}{2}$\\
{\vphantom{1}}\hspace{24.5ex}$\rho_{\a(s)\ad(s-1)},~\delta_g \rho_{\a(s)\ad(s-1)}=0,~[\rho_{\a(s)\ad(s-1)}]=\tfrac{3}{2}$
\end{enumerate}
The second solution corresponds to a formulation with $8s^2+4~\times~8s^2+4$ d.o.f. 
based on the superfields $H_{\a(s)\ad(s)}$~,~$\chi_{\a(s-1)\ad(s-2)}$. The off-shell 
auxiliary spectrum has:
\begin{enumerate}
\item[1.] one real $(s, \, s)$-tensor : $A_{\a(s)\ad(s)},~A_{\a(s)\ad(s)}=\bar{A}_{\a(s)\ad(s)},~\delta_g A_{\a(s)\ad(s)}=0,~[A_{\a(s)\ad(s)}]=2$\\
\item[2.] two real $(s-1,s-1)$-tensors :\\
$u_{\a(s-1)\ad(s-1)},~u_{\a(s-1)\ad(s-1)}=\bar{u}_{\a(s-1)\ad(s-1)},~\delta_g u_{\a(s-1)\ad(s-1)}=0,~[u_{\a(s-1)\ad(s-1)}]=2$\\
$v_{\a(s-1)\ad(s-1)},~v_{\a(s-1)\ad(s-1)}=\bar{v}_{\a(s-1)\ad(s-1)},~\delta_g v_{\a(s-1)\ad(s-1)}=0,~[v_{\a(s-1)\ad(s-1)}]=2$
\item[3.] two real $(s-2,s-2)$-tensors :\\
$S_{\a(s-1)\ad(s-1)},~S_{\a(s-1)\ad(s-1)}=\bar{S}_{\a(s-1)\ad(s-1)},~\delta_g S_{\a(s-1)\ad(s-1)}=0,~[S_{\a(s-1)\ad(s-1)}]=2$\\
$P_{\a(s-1)\ad(s-1)},~P_{\a(s-1)\ad(s-1)}=\bar{P}_{\a(s-1)\ad(s-1)},~\delta_g P_{\a(s-1)\ad(s-1)}=0,~[P_{\a(s-1)\ad(s-1)}]=2$
\item[4.] one $(s, \, s-2)$-tensor : $U_{\a(s)\ad(s-2)},~\delta_g U_{\a(s)\ad(s-2)}=0,~[U_{\a(s)\ad(s-2)}]=2$
\item[5.] two $(s, \, s-1)$-tensors : $\b_{\a(s)\ad(s-1)},~\delta_g \b_{\a(s)\ad(s-1)}=0,~[\b_{\a(s)\ad(s-1)}]=\tfrac{5}{2}$\\
{\vphantom{1}}\hspace{24.5ex}$\rho_{\a(s)\ad(s-1)},~\delta_g \rho_{\a(s)\ad(s-1)}=0,~[\rho_{\a(s)\ad(s-1)}]=\tfrac{3}{2}$
\end{enumerate}
4 - Gauge transformations.\\
Having the detailed spectrum for each theory, we can deduce the gauge transformations 
for each superfield. For the first solution, the components $\chi^{(1,0)(S)}$, $\chi^{(1,0)
(A)}$, $\chi^{(1,2)(S)}$, $\chi^{(1,1)(S,S)}$, $\chi^{(1,1)(S,A)}$ must be gauged away 
while the components $\chi^{(1,2)(A)}$, $\chi^{(2,1)(S)}$, $\chi^{(2,1)(A)}$, $\chi^{(1,1)
(A,A)}$ must not be removed. Using Tables 2,3 and demand a non zero overlap with 
the gauge parameter of $H_{\a(s)\ad(s)}$ we deduce that
\bea
\delta_g\chi_{\a(s)\ad(s-1)}={\Bar {\rm D}}{}  ^2L_{\a(s)\ad(s-1)}+{\rm D}{}^{\b}\Lambda_{
\b\a(s)\ad(s-1)} ~. 
\eea
Up to overall redefinitions of $\chi$ this is exactly the transformation of the compensator 
in \cite{G1,G2} and therefore it will lead to the construction of the same superspace action.
For the second solution, following similar arguments one can find the proper transformation 
for $\chi_{\a(s-1)\ad(s-2)}$ and it is in agreement with the results in \cite{G1,G2}.
%\bea
%\delta_{g}\chi_{\a(s-1)\ad(s-2)}=&~{\Bar {\rm D}}{}  ^{\ad_{s-1}}{\rm D}{}^{\a_s}L_{\a(s)\ad(s-
%1)}+\tfrac{s-1}{s}{\rm D}{}^{\a_s}{\Bar {\rm D}}{}  ^{\ad_{s-1}}L_{\a(s)\ad(s-1)}\\
%&+\tfrac{1}{(s-2)!}{\Bar {\rm D}}{}  _{(\ad_{s-2}}J_{\a(s-1)\ad(s-3))}
%\eea
%which matches with that in \cite{G1,G2}.
%%%%%%%%%%%%%%%%%%%%%%%%%%%%%%%%%%%%%%
%%%%%%%%%%%%%%%%%%%%%%%%%%%%%%%%%%%%%%
%%%%% - Higher Integer spin - %%%%
\section{Arbitrary Integer superspin}
So far, we have worked out in detail the application of the method to the half-integer superspin 
theories. In this section we will appy it to arbitrary integer superspin $Y=s$.\\
{}\\
%%%
1 - The dynamics of the theory will be described by one spin $s$ and one spin $s+\tfrac{1}{2}$. 
Therefore the off-shell degrees of freedom they provide are
\Ibea{l}
\text{spin}~s : (j^2+2)|_{j=s} = s^2+2~,\\
\text{spin}~s+\tfrac{1}{2} : (4j^2+4j+4)|_{j=s} = 4s^2+4s+4
\Ieea
and they will be described by the components $h_{\a(s)\ad(s)}$, $h_{\a(s-2)\ad(s-2)}$, $\psi_{
\a(s+1)\ad(s)}$, $\psi_{\a(s)\ad(s-1)}$, $\psi_{\a(s-1)\ad(s-2)}$ respectively together with 
appropriate gauge transformations.\\
{}\\
2 - The highest spin is $s+1/2$, therefore the main superfield in the superspace description 
must $\Psi_{\a(s)\ad(s-1)}$ with engineering dimensions of $\tfrac{1}{2}$.\\
{}\\
%%%
3 - Auxiliary superfields\\
Similar to the arbitrary half-integer case, the main superfield can not generate all components 
required for the off-shell description of the spins. In this case, the superfield $\Psi_{\a(s)\ad(s-1)}
$ does not include a $(s-2\, s-2)$-tensor to play the role of $h_{\a(s-2)\ad(s-2)}$. Following 
the arguments of the previous section, the list of auxiliary superfields we must consider is
$\{V_{\a(s-1)\ad(s-1)}$, $Z_{\a(s-2)\ad(s-2)}$, $W_{\a(s-1)\ad(s-3)}\}$. However, 
the component $h_{\a(s-2)\ad(s-2)}$ has to be real and therefore it is obvious that only one 
of these choices can give rise to a real component of this type and that is a real superfield $V_{\a(s-1)\ad(s-1)}$.
Nevertheless, for the sake of 
thoroughness we will entertain the possibility of other options and we will see that the 
Diophantine equation will reject them. 

So let us assume that we have ${\cal N}_1$ copies of a real $(s-1,\, s-1)$ superfield, ${\cal 
N}_2$ copies of a real $(s-2,\, s-2)$ superfield and ${\cal N}_3$ copies of a $(s-1,\, s-3)$ 
superfield. These provide the following set of potential auxiliary components
\footnote{Because all of the superfields are bosonic there are no auxiliary fermions 
and because we are in the case of arbitrary $s$ all the auxiliary bosons must be of 
type (2).}
\begin{center}
\begin{tabular}{l r l c l c}
 & & Tensor & Dimensions & off-shell d.o.f. & multiplicity\\
$1~\times~\Psi_{\a(s)\ad(s-1)}$:~&
Fermions: & $(s, \, s-1)$ & [$\tfrac{5}{2}$] & $2(s+1)s$ & $1$ \\
\vspace{-1.5ex}\\
 & Bosons: & $(s+1,\, s-1)$ & [$2$] & $2(s+2)s$ & $\le1$ \\
 &  & $(s-1,\, s-1)$ & [$2$] & $s^2$ & $\le2$\\
 &  & $(s, \, s)$ & [$2$] & $(s+1)^2$ & $\le2$ \\
 &  & $(s, \, s-2)$ & [$2$] & $2(s+1)(s-1)$ & $\le1$ \\
\vspace{-2ex}\\
\hline\hline
\vspace{-2ex}\\
$\mathcal{N}_1~\times~V_{\a(s-1)\ad(s-1)}$:~&
Bosons: & $(s-1,\, s-1)$ & [$2$] & $s^2$ & ${\cal N}_1$ \\
\vspace{-2ex}\\
\hline\hline
\vspace{-2ex}\\
$\mathcal{N}_2~\times~Z_{\a(s-2)\ad(s-2)}$:~&
Bosons: & $(s-2,\, s-2)$ & [$2$] & $(s-1)^2$ & ${\cal N}_{2}$ \\
\vspace{-2ex}\\
\hline\hline
\vspace{-2ex}\\
$\mathcal{N}_3~\times~W_{\a(s-1)\ad(s-3)}$:~&
Bosons: & $(s-1,\, s-3)$ & [$2$] & $2s(s-2)$ & ${\cal N}_3$
\end{tabular}
\end{center}
and the Diophantine equation that balances the bosons with fermions is
\Ibea{rl}
2(s+2)s{\cal N}^{\Psi}_{(s+1,\, s-1)}+2(s+1)s{\cal N}^{\Psi}_{(s, \, s-2)}+2s(s-2){\cal N}_3~&\\
+(s+1)^2{\cal N}^{\Psi}_{(s, \, s)}+s^2[{\cal N}_1+{\cal N}^{\Psi}_{(s-1,\, s-1)}]+(s-1)^2{\cal 
N}_2~&=7(s+1)s+(s+2)\n\label{int2}
\Ieea
where ${\cal N}^{\Psi}_{(s+1,s-1)}\leq1$, ${\cal N}^{\Psi}_{(s,s-2)}\leq1$, ${\cal N}^{\Psi}_{(s,s)}\leq2$, ${\cal N}^{\Psi}_{(s-1,s-1)}\leq2$.
To solve this equation we can repeat the steps we did in the previous section, using (\ref{convert}). Doing that, we quickly find that only identity ($II$) is relevant in this case, hence there is only one economical solution. However, in order to demonstrate the polynomial approach that was mentioned previously, we will follow the latter in this section.  In the polynomial approach,  we equate the coefficients of different powers of $s$ between the left and right hand sided of the equation.  We get:
\Ibea{l}
2{\cal N}^{\Psi}_{(s+1,\, s-1)}+2{\cal N}^{\Psi}_{(s, \, s-2)}+{\cal N}^{\Psi}_{(s, \, s)}+{\cal N}^{\Psi}_{(s-1,\, s-1)}
+{\cal N}_1+{\cal N}_2+2{\cal N}_3=7~,\n\\
%%%
4{\cal N}^{\Psi}_{(s+1,\, s-1)}+2{\cal N}^{\Psi}_{(s, \, s-2)}+2{\cal N}^{\Psi}_{(s, \, s)}-2{\cal N}_2-4{\cal N}_3=8~,\n\\
%%%
{\cal N}^{\Psi}_{(s, \, s)}+{\cal N}_2=2~.\n
\Ieea
Taking into account the upper bound constraints for the various parameters, we find a unique
economical solution that corresponds to:
\Ibea{ll}
{\cal N}_2=0 ~,~&~ {\cal N}^{\Psi}_{(s+1,\, s-1)}=1~, \n\\
{\cal N}_{3}=0 ~,~&~ {\cal N}^{\Psi}_{(s-1,\, s-1)}=2~, \n\\
{\cal N}_1=1 ~,~&~ {\cal N}^{\Psi}_{(s, \, s)}=2~, \n\\
 &~ {\cal N}^{\Psi}_{(s, \, s-2)}=0~. \n
\Ieea
which confirms our suggestion basaed on the reality of the missing ($s-2, s-2$)-tensor component.
The conclusion is that for arbitrary integer superspin theories there is a unique formulation 
of the theory based on a $(s, \, s-1)$ superfield and a real $(s-1,\, s-1)$ superfield with 
$8s^2+8s+4~\times~8s^2+8s+4$ components.  The off-shell auxiliary component 
spectrum of it has:
\begin{enumerate}
\item[1.] two real $(s, \, s)$-tensors :\\
$u_{\a(s)\ad(s)},~u_{\a(s)\ad(s)}=\bar{u}_{\a(s)\ad(s)},~\delta_g u_{\a(s)\ad(s)}=0,~[u_{\a(s)\ad(s)}]=2$\\
$v_{\a(s)\ad(s)},~v_{\a(s)\ad(s)}=\bar{v}_{\a(s)\ad(s)},~\delta_g v_{\a(s)\ad(s)}=0,~[v_{\a(s)\ad(s)}]=2$
\item[2.] three real $(s-1,s-1)$-tensors :\\
$A_{\a(s-1)\ad(s-1)},~A_{\a(s-1)\ad(s-1)}=\bar{A}_{\a(s-1)\ad(s-1)},~\delta_g A_{\a(s-1)\ad(s-1)}=0,~[A_{\a(s-1)\ad(s-1)}]=2$\\
$S_{\a(s-1)\ad(s-1)},~S_{\a(s-1)\ad(s-1)}=\bar{S}_{\a(s-1)\ad(s-1)},~\delta_g S_{\a(s-1)\ad(s-1)}=0,~[S_{\a(s-1)\ad(s-1)}]=2$\\
$P_{\a(s-1)\ad(s-1)},~P_{\a(s-1)\ad(s-1)}=\bar{P}_{\a(s-1)\ad(s-1)},~\delta_g P_{\a(s-1)\ad(s-1)}=0,~[P_{\a(s-1)\ad(s-1)}]=2$
\item[3.] one $(s+1, \, s-1)$-tensor : $U_{\a(s+1)\ad(s-1)},~\delta_g U_{\a(s+1)\ad(s-1)}=0,~[U_{\a(s+1)\ad(s-1)}]=2$
\end{enumerate}
%%%%
4 - To find the gauge transformation we need to have for the superfields $\Psi_{\a(s)\ad(s-1)}
$ and $V_{\a(s-1)\ad(s-1)}$, we go through the list of components that must be removed (like 
$\Psi^{(2,1)(A)}_{\a(s)\ad(s-2)}$) and via Tables 2 and 3 select the appropriate transformation 
that does it while preserving the components that must survive (like $V^{(1,1)(A,A)}_{\a(s-2)
\ad(s-2)}$). It is straight forward to find that the transformations we must have are:
\bea  
\delta_g\Psi_{\a(s)\ad(s-1)}={\rm D}{}^2L_{\a(s)\ad(s-1)}+\tfrac{1}{(s-1)!}{\Bar {\rm D}}{}  _{(
\ad_{s-1}}\Lambda_{\a(s)\ad(s-2))}    \\
%%$
\delta_g V_{\a(s-1)\ad(s-1)}={\rm D}{}^{\a_s}L_{\a(s)\ad(s-1)}+{\Bar {\rm D}}{}  ^{\ad_s}\bar{
L}_{\a(s-1)\ad(s)} ~.   
\eea
Up to redefinitions, these transformations match exactly the ones in \cite{G1,G3} and 
therefore will lead to the construction of the same superspace action.
%%%%%%%%%%%%%%%%%%%%%%%%%%%%%%%%%%%%%%
%%%%%%%%%%%%%%%%%%%%%%%%%%%%%%%%%%%%%%
%%%%%%%%% - Discussion - %%%%%%%%%%%%%
\section{Discussion}
To summarize our results, we have shown that
under the assumption of the natural
requirments of (\emph{i}) \emph{Supersymmetry} [equality of bosonic and fermionic degrees of freedom] and
(\emph{ii}) \emph{Superspace} [all fields must be generated out of superfields], the problem of off-shell completion of 
higher spin supermultiplets [finding the list of required supersymmetric auxiliary fields starting from the on-shell data] can be reduced to a set of Diophantine equations. An top of that if we assume
(\emph{iii}) \emph{Economy} [having no more than what is required], we get a handful of solutions
that correspond to the off-shell spectrum of known irreducible higher spin supermultiplets. 
This new approach provides:
\begin{enumerate}
\item[1.] a method of classifying all irreducible formulations of a
free, massless, arbitrary spin supersymmetric theory
\footnote{Because of the \emph{Supersymmetry} requirement,
the list of auxiliary fields of a supersymmetric theory must correspond to a solution of the Diophantine equation. Hence a classification of the acceptable solutions is a classification of all the possible formulations of the various supermultiplets}
,
\item[2.] a very natural explanation for why some supermultiplets have more than one formulations (\emph{e.g.} the matter-gravitino, supergravity and half-integer superspin supermultiplets) and others do not (integer spin supermultiplet),
\item[3.] a methodology which gives the explicit off-shell component spectrum for a supermultiplet without knowing the action.
The superspace action and the superspace gauge transformation laws can easily be constructed as a direct by-product of this analysis.
\end{enumerate}
Furthermore, for any free, massless, $4D,~{\cal N}=1$ theory 
the supersymmetric auxiliary fields are extremely constrained.
Our analysis provides a very good understanding for why that is.
We have proved that:
\begin{enumerate}
\item[\emph{i}.] The fermionic auxiliary fields must always come in pairs $(\b_{\dots},\rho_{\dots})$ and they appear in the action through algebraic terms of the form $\beta^{\dots}\rho_{\dots}+c.c.$~. Also, they are gauge invariant [$\delta_g \b_{\dots}=\delta_g \rho_{\dots}=0$] and will exist only
if in the superspace description of the theory there is a fermionic superfield with engineering dimensions $\tfrac{1}{2}$.
\item[\emph{ii}.] The bosonic auxiliary fields come in two types called Type (1) and Type (2).\\
Type (1) fields always come in pairs $(A_{\a\ad},B_{\a\b})$
of a ($1,1$)-tensor $A_{\a\ad}$, with $A_{\a\ad}=\pm\bar{A}_{\a\ad}$ and a ($2,0$)-tensor $B_{\a\b}$ with $B_{\a\b}=B_{\b\a}$. They appear in the action through terms of the form $B^{\a\b}\pa_{\b}{}^{\ad}A_{\a\ad}+c.c.$ and they have non-trivial gauge transformations
\Ibea{l}
\delta_g A_{\a\ad}=\pa_{\a\ad}\lambda~,~\lambda=\pm\bar{\lambda}~,~~
\delta_g B_{\a\b}=\pa_{(\a}{}^{\bd}\ell_{\b)\bd}~,~\ell_{\b\bd}=\mp\bar{\ell}_{\b\bd}~,~
\ell_{\b\bd}\sim\ell_{\b\bd}+\pa_{\b\bd}\ell~,~\ell=\mp\bar{\ell}~.
\Ieea
\item[\emph{iii}.] Type (2) auxiliary fields come in singlets $(A_{\dots})$, are gauge invariant [$\delta_g A_{\dots}=0$] and appear in the action through the algebraic terms $A^{\dots}A_{\dots}$.
\end{enumerate}

These Type (1) auxiliary fields, because of their specific index structure, appear only in low spin supersymmetric theories ($j\le2$)
and they are the reason why for low spins there is a zoo of different formulations of the various supermultiplets. The Diophantine equations
generated for these systems $Y=1$ (\ref{D1}) and $Y=\tfrac{3}{2}$ (\ref{msg}) illustrate in a very clear way the interplay between the Type (1) and Type (2) fields which allowed  more solutions, thus more than one superspace formulations.
 
On the other hand for higher spin supermultiplets all auxiliary bosons are of Type (2) in agreement with the results \cite{K1,K2,G1,G2,G3}.
Nevertheless, the Diophantine point of view through (\ref{convert}) provides a clear picture of how one can balance the bosonic and fermionic auxiliary d.o.f. Now we understand that for half-integer spins one can solve the equation (\ref{int-s-d}) in two different ways, using identities ($I$) and ($II$), leading to the two different superspace formulations of $Y=s+\tfrac{1}{2}$ supermultiplets. In contrast, for integer spins and 
equation (\ref{int2}) only identity ($II$) is relevant, hence we get a single solution which corresponds to the one superspace formulation of $Y=s$ supermultiplets.

The results presented in this work depend in a very particular manner of embedding the component fields
into supermultiplets. This embedding was provided by our assumption for the existence of an underlying superspace
formulation that makes the symmetry manifest. An interesting alternative approach to this embedding is the method
of ``\emph{adinkras}'' \cite{adnk1} which provide a one dimensional network graph description of supermultiplets.  
In \cite{adnkMAC,adnkD4N4YM} examples are given for how such an embedding works. The adinkras are being used in order to
generate a set of adjacency matrices associated with them which provide the various representations of suppersymmetry.
These adjacency matrices satisfy a set of algebraic relations called ``Garden Algebra''\cite{GA1,GA2,GA3,GA4}. In \cite{adnkMAC} it was shown that
one can use the Garden algebra in order to find off-shell completions of supersymmetric theories, by generating a system of
{\em {quadratic}} equations. The spirit of the work presented here is very similar to that and explores similar if not the same issues,
where the role of the Garden algebra and its consequence is played by the set of Diophantine equations.

\newpage
%%%%%%%%%%%%%%%%%%%%%%%%%%%%%%%%%%
%%%%%%%%%%%%%%%%%%%%%%%%%%%%%%%%%%
%%%%%%%%%%%%%%%%%%%%%%%%%%%%%%%%%%
 \vspace{.05in}
 \begin{center}
\parbox{4in}{{\it ``Perhaps the topic [of this book] will appear fairly difficult 
to you because it is not yet familiar knowledge and the understanding of 
beginners is easily confused by mistakes; but with your inspiration and 
my teaching it will be easy for you to master, because clear intelligence 
supported by good lessons is a fast route to knowledge.'' \\ ${~}$ 
 \\ ${~}$ 
\\ ${~}$ }\,\,-\,\, Diophantus $~~~~~~~~~$}
 \parbox{4in}{
 $~~$}  
 \end{center}

 \noindent
{\bf Acknowledgements}\\[.1in] \indent
This work was partially supported by the National Science Foundation grant 
PHY-1315155.  This research was also supported in part by CSPT. The work of K.\ K.\ was
supported by the grant P201/12/G028 of the Grant agency of Czech Republic. K.\ K.\ acknowledges the warm hospitality of the Physics departmens
of the University of Maryland and Brown University.  S.\ J.\ G.\ acknowledges 
the generous support of the Provostial Visiting Professorship Program and 
the Department of Physics at Brown University for the very congenial and 
generous hospitality during the period of this work.  

%%%%%%%%%%%%%%%%%%%%%%%%%%%%%%%%%%
%%%%%%%%%%%%%%%%%%%%%%%%%%%%%%%%%%
%%%%%%%%%%%%%%%%%%%%%%%%%%%%%%%%%%
$$~~$$

\end{document}